\documentclass[preprint,12pt]{elsarticle}
\usepackage{array}
\usepackage{booktabs}
\usepackage{multirow}
\usepackage{adjustbox}
\usepackage{color}
\usepackage{xcolor}
\usepackage{graphicx}
\usepackage{hyperref}
\usepackage{makecell}
\usepackage{array}
\usepackage{comment}
\usepackage{amssymb}
\usepackage{amsmath}
\usepackage{xurl}
\usepackage{url}

\begin{document}
\begin{frontmatter}

\title{Capsule Network–Based Multimodal Fusion for Mortgage Risk Assessment from Unstructured Data Sources}

\author[western]{Mahsa Tavakoli}
\author[unsw]{Rohitash Chandra}
\author[western]{Cristián Bravo}

\address[western]{Department of Statistical and Actuarial Science, 
University of Western Ontario, London, Ontario, Canada, N6A 3K7. 
Email: \texttt{mtavako5@uwo.ca}, \texttt{cbravoro@uwo.ca}}
\address[unsw]{Transitional Artificial Intelligence Research Group, 
School of Mathematics and Statistics, University of New South Wales, 
Sydney, Australia. 
Email: \texttt{rohitash.chandra@unsw.edu.au}}

\begin{abstract}
Mortgage risk assessment traditionally relies on structured financial data that is proprietary and often inaccessible to the public. This limits the ability of smaller institutions, advisors, and individuals to evaluate mortgage-related risks using cost-free information sources. Meanwhile, advances in deep learning and the availability of unstructured data---such as financial news, sentiment cues, and spatial imagery---create opportunities to approximate risk assessments without depending on internal credit datasets.

This study introduces a multimodal framework that integrates diverse unstructured signals to predict mortgage default risk. At its core is FusionCapsNet, a capsule-inspired fusion architecture that extends beyond a single capsule layer by restructuring how heterogeneous modalities are processed and integrated. Rather than compressing all features into a common vector, FusionCapsNet expands modality-specific encodings into multiple capsule projections, enabling routing-by-agreement within each modality's latent embedding space while preserving semantic and contextual relationships that are often weakened in traditional fusion methods. A dynamic weighting mechanism further adapts the contribution of each modality by aggregating modality-level evidence summarized by normalized confidence scores, allowing the gating network to learn instance-specific modality contributions while integrating complementary information across modalities.

The framework integrates sentiment extracted from financial news, LiDAR-based spatial features, and textual embeddings within an interpretable fusion strategy supported by Grad-CAM visualization. Empirically, FusionCapsNet delivers an improvement over benchmark fusion approaches, increasing the AUC by approximately 3 percentage points relative to the strongest baseline. These results highlight the potential of publicly available unstructured data, combined with capsule-inspired fusion, to support cost-free and explainable mortgage risk assessment.
\end{abstract}

\begin{keyword}
Deep learning, Mortgage Risk, Sentiment Analysis, Information Fusion, Multimodality, Capsule Networks
\end{keyword}
\end{frontmatter}

\section{Introduction}
\label{sec:Introduction}

Exploring the factors that influence mortgage defaults has significance in finance and risk management\cite{bhattacharya2019bayesian}. For example, investing in the mortgage market during periods of high defaults can lead to substantial drops in asset values, potentially leading to wealth erosion\cite{bhattacharya2019bayesian}. Furthermore, high default rates can strain financial institutions \cite{ntiamoah2014loan}, impacting their lending capabilities and overall financial stability. High default rates can lead to tighter credit markets, making it harder for individuals and businesses to secure loans, further affecting economic growth and stability. Understanding these risks is therefore crucial for investors, policymakers, and financial institutions to mitigate possible negative impacts at both individual and broader economic levels. At the same time, mortgage-default risk is not only a concern for large banks and institutional investors. Small businesses, independent financial advisors, portfolio managers, and even individual investors increasingly need timely indicators of regional mortgage risk to guide asset allocation, pricing, and regulatory decisions. However, most operational credit-risk systems are based on proprietary scorecards and structured loan data that are costly, confidential, and unavailable to these stakeholders\cite{dastile2020statistical}. This creates a gap between the practical importance of mortgage-default prediction and the accessibility of the data on which current models are built.

A large body of research has focused on the prediction of mortgage defaults using structured data sets\cite{bhattacharya2019bayesian}, particularly loan characteristics and the demographics of the borrower\cite{ntiamoah2014loan}. These studies consistently show that such variables strongly influence the lenders’ decisions when assessing credit risk \cite{ozturkkal2024explaining}. More broadly, statistical and machine-learning models for credit scoring and default prediction have been extensively developed for structured numerical features such as balances, payment histories, loan-to-value ratios, and borrower income \cite{dastile2020statistical,jiang2018loan}. These approaches have led to mature practices for model development, validation, and governance, but fundamentally assume privileged access to detailed financial records.

Market participants outside of major institutions may also need to understand the risks of mortgage default using the information available to them. Access to structured financial data is often challenging because such data is typically controlled by large institutions and restricted due to privacy concerns. This limitation reduces public access and creates a significant barrier to obtaining critical information. Publicly available data is therefore highly valuable, as they can provide insights into mortgage risk without relying on proprietary financial datasets. Leveraging such data allows market participants with limited resources to make more informed mortgage investment decisions and enables policymakers to design evidence-based legislative or regulatory measures.

In the field of data analytics, structured data has long played a key role in supporting predictive modeling tasks \cite{dastile2020statistical}. However, the potential of unstructured data from multimedia sources, such as text and images, remains largely untapped and can provide valuable and often unexpected insights \cite{zhang2020combining}.  The volume of unstructured and semi-structured data continues to increase with the rapid growth of big data, drawing significant attention from researchers and practitioners. In the realm of mortgage default prediction, structured data has been extensively studied \cite{jiang2018loan}, while the integration of unstructured data remains in its infancy due to challenges such as data preprocessing, feature extraction, and model interpretability. Despite these obstacles, unstructured data offers the potential to enrich feature representations and enhance predictive accuracy \cite{stevenson2021value}. Therefore, two publicly available sources are particularly relevant for mortgage risk within this broader unstructured-data landscape, including geospatial imagery and financial news.

Unstructured data sources that are openly accessible and readily available include materials such as news articles and remote-sensing images. These sources offer valuable analytical content, providing opportunities to extract meaningful insights \cite{adrian2021sentinel}. The \textit{Earth Engine Data Catalog} \cite{earthenginecatalog}
 hosts LiDAR (Light Detection and Ranging) data as part of its openly accessible collections through \textit{Google Earth Engine}\cite{google_earth_engine}
. Combining spatial data, such as elevation maps, with text or financial indicators can substantially enhance predictive modeling. For example, Stevenson et al.\ \cite{stevenson2022deep} showed that LiDAR-generated elevation maps can be transformed into meaningful embeddings to model socioeconomic indicators. 

Financial news media is another relevant unstructured data source, which plays a critical role in shaping investment decisions, market sentiment, and risk perceptions \cite{feuerriegel2019news}. The mortgage market is highly sensitive to economic conditions, policy changes, and macro-financial developments \cite{zhu2017housing,iselin2016news}. Negative news on economic stability can raise the perceived risk of default, while positive news can reduce it. By incorporating news into predictive models, lenders and analysts can proactively identify risks and adjust decision-making accordingly. Although prior research has explored unstructured financial reports \cite{tavakoli2025multi}, loan application narratives \cite{xia2020predicting}, and social media activity \cite{ge2017predicting}, the combined use of spatial map features and financial news for mortgage-risk modeling remains largely unexplored.

Integrating heterogeneous unstructured data requires multimodal techniques capable of capturing complementary signals in different formats \cite{zhao2024deep}. Multimodal models have gained attention due to their ability to integrate structured and unstructured sources\cite{zheng2019multimodal}. However, their performance is strongly influenced by the fusion strategy employed \cite{pawlowski2023effective}. Simple fusion approaches, such as concatenation or addition, can compress modality-specific signals without explicitly preserving their internal structure \cite{tavakoli2025multi}. Capsule Networks \cite{sabour2017dynamic} offer an alternative by representing features as vectors and using routing-by-agreement to preserve part–whole relationships, but existing capsule architectures were designed for single-modality tasks and do not directly address multimodal fusion.

Our study aims to examine the relationship between selected publicly available unstructured features and mortgage default risk, demonstrating how these features can serve as practical indicators for broader market participants. 
To achieve this, we develop a novel deep learning framework, \textit{FusionCapsNet}, which extends beyond traditional fusion techniques by preserving structured, contextual, and modality-specific representations in latent embedding space, while adaptively weighting each modality to enhance predictive performance and interpretability. Routing-by-agreement is performed independently within each modality; cross-modal integration occurs subsequently through the adaptive gating mechanism.

Our case study uses openly accessible unstructured data streams, including remote-sensing imagery and categorized financial news documents, combined with numerical features derived from sentiment. FusionCapsNet can integrate diverse data sources and demonstrate that public unstructured data can approximate commercial mortgage-risk assessments, without access to proprietary loan-level records.

The core strengths of our study are:
\begin{itemize}
    \item Capturing multiple perspectives while preserving the diversity of each source.
    \item Strengthening complementary modality-level evidence through adaptive aggregation while filtering noise.
    \item Adaptive weighting of information based on modality-specific relevance.
    \item Ensuring interpretability and accessibility for users who lack proprietary data.
\end{itemize}

In summary, we contribute (i) a problem formulation in which mortgage-default risk is predicted using only publicly available unstructured data; (ii) a capsule-based multimodal fusion architecture tailored to heterogeneous text, image and sentiment channels; and (iii) an empirical comparison showing that this framework outperforms standard fusion baselines while retaining interpretability.

The structure of this paper is as follows. Section~\ref{sec:background} offers a review of the relevant literature, highlighting the unique contributions of this study. In Section~\ref{sec:methodology}, we elaborate on the methodology employed, including details of the dataset and the channels utilized in the study. We provide the results in Section~\ref{sec:result}. Finally, Section~\ref{sec:conclusion} highlights the contributions and concludes the key findings.

\section{Background and Related Work}
\label{sec:background}

\subsection{Structured-Data Modeling in Mortgage Default Prediction}

Mortgage-default prediction traditionally relies on numerical datasets that contain quantitative variables related to loans and borrower characteristics~\cite{kealhofer2003quantifying,gao2023severe,zheng2023community,nwafor2023determinants}. These datasets typically include features such as loan amounts, interest rates, credit scores, income levels, debt-to-income and loan-to-value ratios, payment histories, property values and other financial indicators \cite{cowden2019default,siering2023peer,xia2020predicting,saavedra2024probability}. Although these structured datasets support mature risk-modeling practices, they may require privileged access to sensitive and proprietary financial records.

A large body of prior work has focused on the prediction of mortgage defaults using structured numerical data, including loan attributes and demographics of the borrower \cite{bhattacharya2019bayesian,ntiamoah2014loan}. Machine-learning models such as logistic regression, tree-based methods, and deep neural networks effectively leverage structured financial indicators \cite{dastile2020statistical,jiang2018loan}. These methods consistently demonstrate strong predictive performance, but depend on detailed financial datasets that are costly and unavailable to many public stakeholders. In this work, we extend this view by providing a holistic view of credit risk using a combination of publicly available data in both structured and unstructured form. 

\subsection{Unstructured-Data Analytics for Socioeconomic and Financial Prediction}

Although structured datasets are widely used, the increasing availability of unstructured data—text, imagery, and geospatial sources—offers alternative avenues for inferring socioeconomic and financial patterns \cite{zhang2020combining,chen2014big}. We find that two categories are most relevant for the assessment of mortgage-risk in this space, which consists of spatial imagery and financial news.

Remote sensing technologies such as LiDAR, satellite imagery, and Earth Engine mapping have been used extensively to study socioeconomic and environmental phenomena \cite{lu2013remote,grove2014ecology,shanahan2014socio}. These spatial data sources provide fine-grained geographic information and are cost-effective alternatives in regions with limited census or administrative data \cite{chen2002approach,li2007measuring}. Machine learning applications have demonstrated the value of spatial imagery for socioeconomic inference. Block et al.~\cite{block2017unsupervised} and Jean et al.~\cite{jean2016combining} used high-resolution satellite images to estimate poverty and development indicators, using features such as luminosity, vegetation and land cover. Law et al.~\cite{law2019take} improved house-price prediction using visual features extracted from satellite and street-view images. Zou et al.~\cite{zou2021detecting} developed a deep model to detect abandoned houses from satellite imagery. Suel et al.~\cite{suel2021multimodal} used deep learning with imagery to predict deprivation indices.

Additional studies have applied satellite and LiDAR imagery to environmental sciences, urban-growth analysis, and urban form study \cite{pan2020land,hamraz2019deep,zhou2020LiDAR}. Lu et al.~\cite{lu2011volumetric} and Warth et al.~\cite{warth2020prediction} demonstrated that LiDAR captures urban characteristics—such as building shapes, green spaces, and street configuration—associated with socioeconomic outcomes. In general, these works highlight the potential of LiDAR-derived representations to support the assessment of mortgage risk by capturing neighborhood quality, environmental context, and socioeconomic indicators \cite{li2007measuring}. Following this body of work, we are including LiDAR-based data in our analysis as the main image-based source of data.

A secondary source of unstructured data, textual information in the form of financial news, has become increasingly important for understanding market sentiment and economic risk \cite{li2017web}. Deep learning has advanced news classification and sentiment analysis, allowing automated extraction of financial signals \cite{farimani2022investigating,chen2022comparative,minaee2021deep,nasir2021fake}.

Transformer-based models \cite{vaswani2017attention}, particularly Bidirectional Encoder Representation from Transformers (BERT) \cite{devlin2018bert}  capture contextual relationships across full sequences and have become central to NLP tasks such as sentiment analysis, classification, and event detection. Considerable research has analyzed financial news using BERT-based models, while modern large language models rely on their decoder architecture.

Recent studies highlight the value of text-based analysis in financial modeling. Liao et al.~\cite{liao2021integrated} introduced a multitask learning framework for the detection of fake-news. Hu et al.~\cite{hu2022causal} proposed a causal-inference-based model to mitigate image-text matching bias in multimodal fake-news detection. Li et al.~\cite{li2020multimodal} examined multimodal stock prediction using event-driven LSTMs. Other work integrates visual and textual information in news articles to improve event detection and geolocation \cite{ramisa2017breakingnews,yang2019shared}.

Despite substantial progress in textual news analytics, the application of financial news to the prediction of mortgage default remains limited, so in this work we have collected an extensive dataset of news that will form an integral part of our multimodal system for mortgage default prediction.

\subsection{Multimodal Data Fusion}

As heterogeneous datasets become more common, multimodal learning has gained prominence in financial applications. Todd et al.~\cite{todd2025multimodal} fused text with vocal features from earnings calls to improve sentiment classification. Wang et al.~\cite{Wang2023attentive} combined textual information with financial ratios to detect statement fraud.

Many multimodal architectures rely on simple fusion mechanisms such as concatenation, addition, or multiplication \cite{jiao2024comprehensive,peng2025multimodal,Gaonkar2021comprehensive}. Although computationally efficient, these strategies often compress modality-specific signals and lose fine-grained cross-modal relationships. More advanced methods, including cross-attention and transformer-based fusion, address some of these limitations \cite{gan2024multimodal,wang2025cross,an2024attention,li2024crossfuse}. Recent surveys emphasize a shift toward integrated, attention-based multimodal frameworks, which generally outperform early and late fusion techniques \cite{zhao2024deep}.

However, even attention-based fusion can struggle to preserve structured and modality-specific relationships when integrating heterogeneous data such as images, text, and sentiment. As such, we design a capsule-inspired architecture that combines intra-modal agreement-based routing with subsequent adaptive multimodal fusion.

To conclude this review of the literature and the diagnosis of the identified gap, we summarize our findings. So far, LiDAR information has been shown to be useful for financial analysis, while textual information in the form of news has been shown to be relevant to the financial market as a whole, but it has never been studied as a source of data for mortgage credit risk. In order to be able to properly use these data sources and to extend the literature on multimodal data fusion, we propose a system that is able to take advantage of individual modalities by combining them in a complex view that goes beyond traditional fusion methods. We will develop this system in the next section.

\section{Methodology}
\label{sec:methodology}

\subsection{Capsule Networks}

Sabour et al. \cite{sabour2017dynamic} introduced Capsule Networks that improved traditional CNNs by encoding features as vectors (capsules) rather than scalars, thus preserving both the presence and spatial relationships of features. This vectorized representation, combined with a dynamic routing mechanism, allows capsules to selectively pass information to higher-level capsules based on agreement, effectively addressing limitations of CNNs such as loss of pose and viewpoint information. Capsule networks have shown strong performance in various domains, including medical image analysis \cite{lalonde2021capsules}, hyperspectral image classification \cite{paoletti2018capsule}, and text classification \cite{kim2020text}, often achieving competitive accuracy with fewer parameters. Its ability to retain part–whole relationships makes it particularly suitable for tasks requiring interpretability and robustness to spatial transformations. However,  capsule networks were developed when multimodal data fusion had less interest; therefore, their potential as a fusion method has not been explored.

\subsection{Language Models}  
BERT \cite{devlin2018bert} is a transformer-based language model that sets a benchmark in NLP. Its key innovation lies in bidirectional training, which enables the model to consider both preceding and succeeding words when interpreting a given token. This approach allows BERT to capture complex contextual relationships and dependencies, resulting in a deeper semantic understanding. After pre-training, BERT can be fine-tuned for various downstream tasks such as sentiment analysis \cite{SHUKLA2023100025}, topic modeling \cite{li2024geotpe}, and named entity recognition \cite{liu2023naming}.

BERT has had several other developments, which we describe for completeness. RoBERTa \cite{liu2019roberta} builds on BERT by refining its pretraining methodology, utilizing masked language modeling with dynamic masking, and training on substantially larger datasets with improved optimization strategies. These enhancements enable RoBERTa to achieve superior performance in many NLP tasks. Like BERT, RoBERTa can be fine-tuned in domain-specific datasets, making it adaptable to diverse applications such as named entity recognition \cite{wu2021research} and text classification \cite{briskilal2022ensemble}.
ALBERT \cite{lan2019albert} (A Lite BERT) addresses the computational and memory limitations of BERT by introducing parameter sharing between layers and factorizing the embedding matrix into smaller components. This design reduces the size of the model without sacrificing performance, allowing faster training and inference. ALBERT has shown effectiveness in tasks including text classification \cite{vaca2024interpretability}, sentiment analysis \cite{duan2024hybrid}, and named entity recognition \cite{shishehgarkhaneh2024transformer}.
CNN is a traditional deep learning architecture originally designed for computer vision, but also effective in certain NLP tasks. By applying convolutional filters over text embeddings, CNNs can capture local n-gram features efficiently, making them useful for applications like text classification and sentiment analysis. Despite being simpler than transformer-based models, CNNs can still deliver competitive performance in specific contexts\cite{thekkekara2024attention}, which motivates their inclusion as one of the candidate models in our evaluation.

\subsection{Image models}
VGG19 \cite{bansal2023transfer}, developed by the Visual Geometry Group (VGG), is a classic deep convolutional neural network renowned for its simplicity and strong performance in image classification tasks. The architecture comprises 19 layers, including 16 convolutional layers followed by fully connected layers, enabling a progressive learning process from low-level features such as edges and textures to high-level semantic representations. VGG19 has been widely applied in domains such as image classification \cite{bansal2021transfer}. Bansal et al. \cite{bansal2023transfer} used VGG-19 for an image classification application and showed that it performs better compared to other well-known classifiers.
EfficientNet \cite{tan2019efficientnet} is a CNN architecture designed to achieve an optimal balance between model size, computational efficiency, and accuracy. It uses a neural architecture search \cite{elsken2019neural} to automatically identify the most effective combination of network depth, width, and resolution. This approach enables EfficientNet to be fine-tuned on smaller, domain-specific datasets, significantly reducing training time and computational requirements.  

EfficientNet has shown strong performance in a wide range of applications, including image detection \cite{tan2019efficientnet}. Its versatility extends to classification tasks \cite{ferentinos2018deep}, where it has achieved high accuracy levels, in some cases matching the performance of human experts \cite{gulshan2019effnet}.
Another candidate for image model training is ResNet (Residual Network) \cite{he2016deep}, which addresses the challenges of training very deep neural networks by introducing residual connections. These connections allow information to bypass intermediate layers and feed directly into deeper layers, effectively mitigating the vanishing-gradient problem. Beyond its groundbreaking success in image recognition, ResNet has become a foundational backbone for numerous computer vision tasks, including object detection \cite{zhao2019object} and face recognition \cite{bajpai2025ri}.

\subsection{Multimodal Fusion Framework}
\label{sec:framework}

Existing structured-data models achieve strong performance but depend on proprietary datasets unavailable to many stakeholders. Unstructured data—such as LiDAR imagery and financial news—offers publicly accessible signals with demonstrated socioeconomic and financial relevance, yet their integration into mortgage-default prediction remains limited.
Current multimodal fusion approaches often compress heterogeneous modality embeddings into shared representations, potentially weakening modality-specific characteristics and reducing the richness of modality-specific representations that are important for risk assessment. These limitations motivate the development of a multimodal fusion framework capable of leveraging spatially derived image features, financial news text, and sentiment signals while preserving modality-specific representational diversity and adaptively integrating modality-level evidence summarized by normalized confidence scores at the decision stage.

\begin{figure*}[htbp!]
\centering
{\includegraphics[width=5.5in]{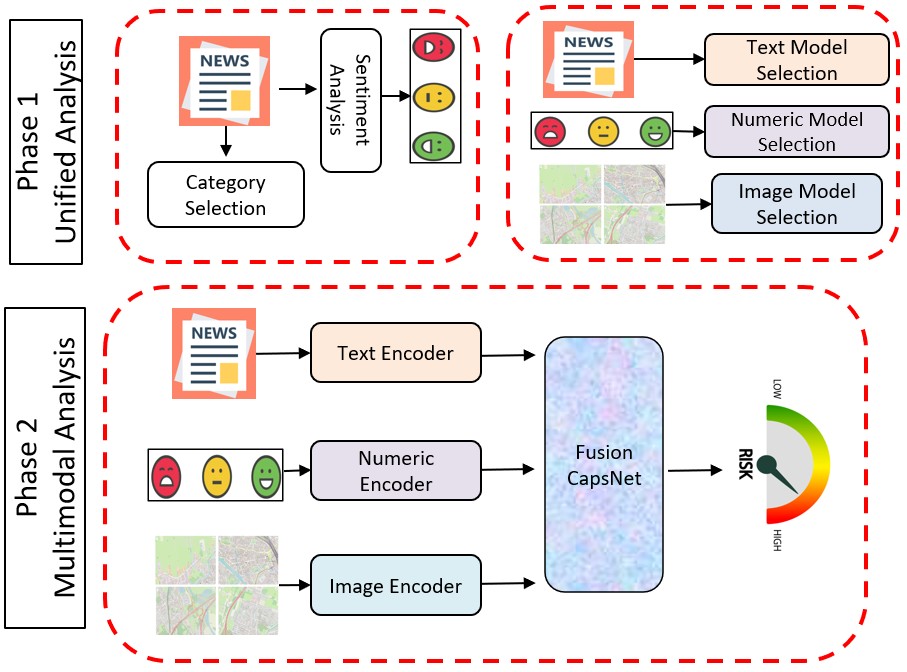}}
\caption{Overview of the framework consisting of two main phases. 
In Phase~1 (\textit{Unified Analysis}), news articles undergo sentiment analysis and category selection, followed by independent model selection for text, numeric sentiment features, and LiDAR images. 
In Phase~2 (\textit{Multimodal Analysis}), the best-performing models serve as modality-specific encoders whose outputs are fused using the proposed FusionCapsNet architecture to predict mortgage default risk.}
\label{fig:framework}
\end{figure*}

Our framework, illustrated in Figure~\ref{fig:framework}, adopts a two-phase approach to mortgage-risk prediction that separates unimodal representation learning from multimodal fusion. This design aims to ensure that each data source contributes a distinct and complementary signal while maintaining transparency over how information flows through the architecture.  

In Phase~1 (unimodal modeling), we independently analyze each modality to identify the most informative and stable representation. We first evaluate LiDAR-derived map imagery using several image-based architectures, and the best-performing network is retained as the image encoder. News articles are then considered across multiple economic categories, yielding a set of text models that capture heterogeneous macro-financial signals. Two news categories exhibit strong and non-redundant predictive capability and are selected as text encoders. In parallel, we aggregate sentiment scores computed from the full news corpus into a separate numeric channel, providing a compact summary of market perception that complements both textual context and image-derived representations.

In Phase~2 (multimodal fusion), we integrate the selected unimodal encoders into a unified architecture known as FusionCapsNet, which is designed to preserve modality-specific structure while adaptively integrating modality-level evidence summarized by normalized confidence scores through a gating mechanism. We process each modality (text, sentiment, and images)  through its own encoder prior to fusion, avoiding early feature mixing and reducing the risk of information loss. The encoded representations are passed to separate modality-specific capsule hierarchies, where dynamic routing aggregates votes for the target class independently within each latent embedding space. Cross-modal integration is introduced only after this intra-modal routing stage, through a gating mechanism that assigns adaptive weights to the modality-level scores. 


This two-phase design provides three complementary benefits within a single methodological framework.

First, interpretability is supported by the modular encoder structure, which processes text, image, and numeric sentiment modalities independently rather than collapsing all signals into a single latent embedding. This design allows each modality to retain its own representation and capsule-based confidence score prior to fusion. As a result, users can explicitly inspect unimodal predictive behavior (Table~\ref{unimodal}), examine modality-specific confidence signals derived from capsule norms and similarity measures (Section~\ref{sec:framework}), and assess each modality’s relative contribution through comparisons between unimodal and fused performance (Tables~\ref{unimodal} and~\ref{performance}). This modularity enables practical interpretability by allowing stakeholders to identify which information source drives a given prediction, rather than relying on opaque joint representations.

Second, information preservation is promoted by delaying multimodal fusion until modality-specific features have been fully extracted. By avoiding early compression of heterogeneous representations, the framework maintains the distinct geometric, semantic, and contextual characteristics of text, image, and sentiment signals. Empirically, this preservation is reflected in the superior discriminative performance of the proposed fusion model relative to early-fusion baselines such as Addition and Concatenation, despite using identical unimodal encoders (Table~\ref{performance}), indicating that meaningful modality-specific information is retained through the fusion process.

Third, adaptive integration is achieved through a gated weighting mechanism that learns instance-specific modality importance in a data-driven manner. The gating network assigns normalized weights to modality-level confidence scores, allowing informative channels to be emphasized while suppressing weaker or noisier inputs. This adaptive behavior contributes to robustness by stabilizing predictions across heterogeneous input conditions and is reflected in the consistent performance gains of the fusion model across repeated experimental runs (Table~\ref{performance}).

The strengths of the FusionCapsNet model emerge directly from the modular structure of the architecture. Since each modality retains its own encoder and confidence score, the contribution of each channel can be independently inspected, providing a transparent form of interpretability. Information preservation arises from delaying fusion until after modality-specific feature extraction, which reduces the risk of representation collapse that occurs in early-fusion approaches. Robustness is supported by the routing-by-agreement mechanism: consistent inter-capsule signals receive higher coupling weights, while noisy or low-agreement representations are naturally suppressed, which is a behavior that has been documented in prior capsule-network research \cite{lalonde2021capsules}.

\paragraph{Clarification on inductive bias.}
Capsule Networks were originally motivated by modeling geometric part--whole relationships and viewpoint invariance in vision tasks.
In the present framework, we do not assume any physical or spatial part--whole relationship between heterogeneous modalities such as financial news semantics and LiDAR-derived elevation maps.
Instead, capsules operate purely in latent embedding space. Routing-by-agreement is therefore interpreted as a structured evidence aggregation mechanism that amplifies directionally consistent latent projections and suppresses noisy or inconsistent signals within each modality-specific latent embedding, rather than modeling geometric pose alignment.

\subsection{FusionCapsNet Framework}

The FusionCapsNet framework is inspired by Capsule Networks (CapsNet) but is not derived from its standard form. Instead, the capsule mechanisms are selectively adapted to each modality, and the fusion process is restructured to accommodate the heterogeneous nature of text, image, and numeric sentiment data. Conventional fusion strategies, such as simple addition or concatenation, typically require that encoder outputs be projected into a common dimensionality before fusion. This compression step can lead to substantial information loss, as rich modality-specific features are forced into a limited, shared representation. In contrast, the CapsNet-inspired fusion mechanism expands each encoded representation into multiple capsules, allowing it to be analyzed from several perspectives. This design helps preserve contextual, and modality-specific details that would otherwise be discarded.

A second motivation for adopting a capsule-based fusion mechanism stems from the intrinsic heterogeneity of the encoded outputs. Text embeddings, numeric sentiment scores, and image-derived features differ in scale, geometry, and semantic interpretation. Forcing them into a uniform numerical format via direct addition or naive concatenation can reduce the richness of modality-specific representations. By processing each modality independently through its own capsule hierarchy, FusionCapsNet preserves modality-specific characteristics while producing modality-level representations summarized by normalized confidence scores that are subsequently integrated through the adaptive gating mechanism.
In comparison with deterministic approaches such as addition and concatenation, which collapse heterogeneous features into a single shared representation, capsule-based fusion retains both the magnitude and orientation of modality-specific vectors, preventing representation collapse. Dynamic routing enables the model to selectively amplify directionally consistent evidence within each modality-specific latent space while suppressing noisy or inconsistent activations. Unlike attention-based fusion, which primarily performs relevance weighting, capsules model structured relationships within each modality's latent embedding space through agreement-based routing, while modality-level information is subsequently integrated through the adaptive gating mechanism. These characteristics allow FusionCapsNet to better exploit complementary signals across modalities, resulting in more stable and discriminative fused representations.

\begin{figure*}[htbp!]
\centering
{\includegraphics[width=5.5in]{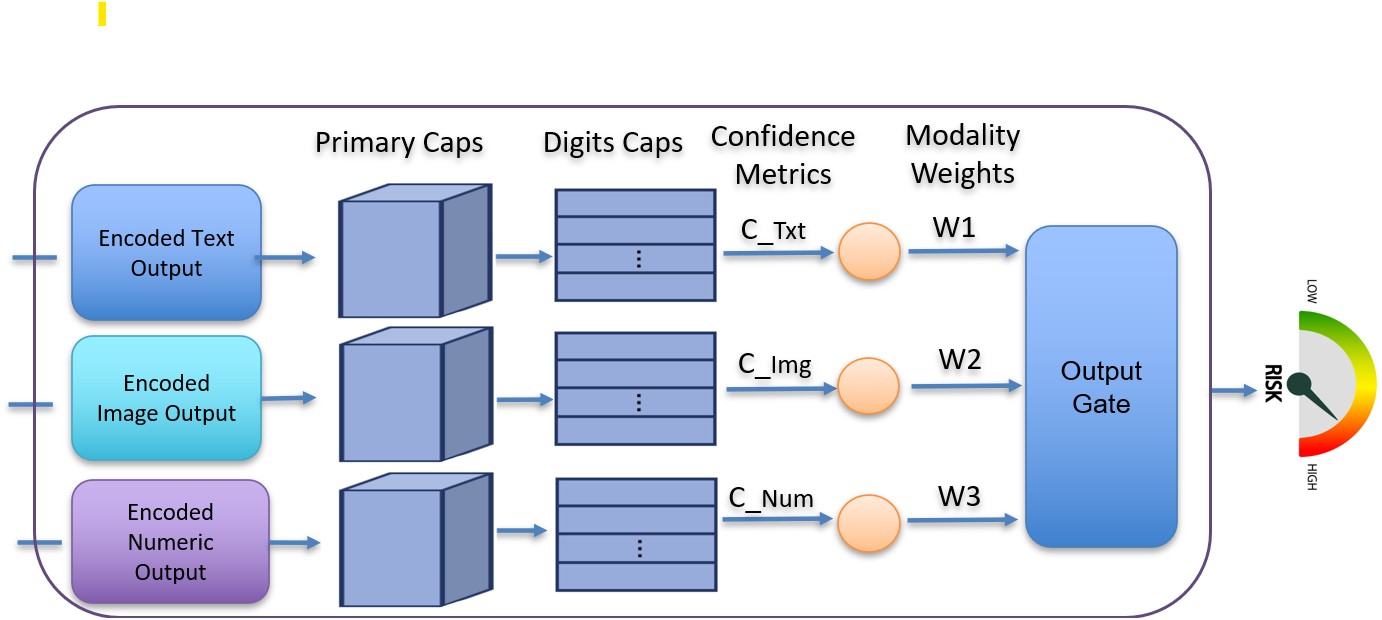}}
\caption{Multimodal capsule-based fusion architecture. 
Encoded outputs from text, image, and numeric modalities are projected into primary capsules and subsequently routed to digit capsules. 
Modality-specific confidence metrics ($C_{\mathrm{Txt}}$, $C_{\mathrm{Img}}$, $C_{\mathrm{Num}}$) are computed and scaled by learnable weights ($W_{1}$, $W_{2}$, $W_{3}$), then passed through an adaptive weighted gate to produce the final risk prediction.}
\label{fig:fusion}
\end{figure*}

Figure~\ref{fig:fusion} illustrates the fusion stage, which integrates the modality-specific encoders selected during Phase~1. For each modality, we first apply the best-performing encoder to the raw input, producing a high-dimensional embedding that retains modality-specific characteristics while providing a compact and discriminative representation. Formally, for modality $i \in \{t,i,n\}$, corresponding to text, image, and numeric sentiment, respectively,
\begin{equation}
\mathbf{E}_i = \mathrm{model}_i(\mathbf{X}_i),
\end{equation}
where $\mathrm{model}_i(\cdot)$ denotes the modality-specific feature extractor and $\mathbf{E}_i$ is the resulting embedding.

\subsubsection{Primary Capsules}

The encoded features of each modality are expanded into a set of primary capsules, where each capsule vector represents a distinct view of the same underlying signal. Rather than compressing features into a single representation, the model projects the embedding of each modality into multiple capsule vectors, enabling it to emphasize different aspects of the modality. For example, in the text channel, one capsule may capture semantic content while another emphasizes sentiment; in the numeric sentiment channel, capsules may reflect intensity or uncertainty; and in the image channel, capsules may highlight structural cues, textures, or neighborhood patterns. This expansion step preserves modality-specific detail while providing richer representations for downstream routing and fusion.

Mathematically, this can be expressed as follows.

Let $B$ denote the batch size, and let $d^{(i)}$ represent the dimensionality of the embedding for modality $i$. For each modality, the encoded feature representation is therefore given by $\mathbf{E}_i \in \mathbb{R}^{B \times d^{(i)}}$.
For each modality $i$, we project its embedding $\mathbf{E}_i \in \mathbb{R}^{B \times d^{(i)}}$ into $N_{\mathrm{pc}}^{(i)}$ primary capsules using a set of projection matrices $\{\mathbf{W}_{\mathrm{pc},k}^{(i)}\}_{k=1}^{N_{\mathrm{pc}}^{(i)}}$, where each $\mathbf{W}_{\mathrm{pc},k}^{(i)} \in \mathbb{R}^{d^{(i)} \times d_{\mathrm{pc}}^{(i)}}$.  

The $k$-th capsule for the modality $i$ is obtained as:  

\begin{equation}
\mathbf{P}_{i,k} = \mathbf{E}_i \mathbf{W}_{\mathrm{pc},k}^{(i)}, \quad \mathbf{P}_{i,k} \in \mathbb{R}^{B \times d_{\mathrm{pc}}^{(i)}}.
\end{equation}

Stacking across all $N_{\mathrm{pc}}^{(i)}$ capsules yields the full set of primary capsules:  

\begin{equation}
\mathbf{P}_i = [\mathbf{P}_{i,1}, \mathbf{P}_{i,2}, \ldots, \mathbf{P}_{i,N_{\mathrm{pc}}^{(i)}}] \in \mathbb{R}^{B \times N_{\mathrm{pc}}^{(i)} \times d_{\mathrm{pc}}^{(i)}}.
\end{equation}

The resulting set of primary capsules $\{\mathbf{P}_{i,k}\}_{k=1}^{N_{\mathrm{pc}}^{(i)}}$ serves as the input to the digit capsule layer, where higher-level representations are formed through transformation and routing by agreement.

\subsubsection{Digit Capsules}

We transform the outputs of the primary capsules and route them into a smaller set of digit capsules, each representing evidence at the class level.
It is important to note that the routing mechanism described below is performed independently within each modality and does not involve direct interaction across modalities.
In the transformation step, each lower-level feature group is reshaped to generate a prediction of what a higher-level feature might look like (e.g., a risk class).
These predictions are then refined through routing by agreement, where signals that consistently align within each modality-specific latent space are amplified, and inconsistent signals are down-weighted. 
As a result, the digit capsules integrate coherent evidence within each modality while suppressing noise. This stage serves three main purposes: 
\begin{enumerate}
   
 \item  Ensuring consistency by reinforcing features that agree across capsules within the same modality;
 
  \item  Preserving structure by combining lower-level elements into whole entities such as “high-risk” or “low-risk” mortgages in our application;
  
   \item  Enabling adaptive weighting, so that irrelevant or noisy capsules contribute less while informative ones dominate the decision-making process.  

\end{enumerate}

Therefore, for each modality $i$:
\begin{equation}
\widehat{\mathbf{u}}_{j|k}^{(i)} = \mathbf{P}_{i,k} \mathbf{W}_{kj}^{(i)}, \quad
\mathbf{S}_j^{(i)} = \sum_{k} c_{kj}^{(i)}\,\widehat{\mathbf{u}}_{j|k}^{(i)},
\end{equation}
where $\mathbf{W}_{kj}^{(i)}$ are the \textit{transformation matrices} 
that project each primary capsule $\mathbf{P}_{i,k}$ into the space of the digit capsule $j$
(note that these are distinct from the projection matrices $\mathbf{W}_{\mathrm{pc},k}^{(i)}$ 
used earlier to form the primary capsules), 
$c_{kj}^{(i)}$ are routing coefficients, 
and $\mathbf{S}_i \in \mathbb{R}^{B\times N_c\times d_{\text{dc}}^{(i)}}$ 
is the output passed to the confidence metric stage.

We specify the routing update used in all the modalities where the coupling coefficients are initialized as $b_{kj}^{(i)}=0$ and updated for $R=3$ routing iterations using
\[
c_{kj}^{(i)}=\frac{\exp(b_{kj}^{(i)})}{\sum_{j'}\exp(b_{kj'}^{(i)})}, \qquad
b_{kj}^{(i)} \leftarrow b_{kj}^{(i)} + \widehat{\mathbf{u}}_{j|k}^{(i)} \cdot \mathbf{S}_j^{(i)}.
\]
This follows the standard dynamic-routing procedure of Sabour et al. \cite{sabour2017dynamic}, ensuring reproducibility.

From a latent-space perspective, dynamic routing can be interpreted as an iterative consistency-maximization process.
The coupling coefficients $c_{kj}^{(i)}$ are updated to increase alignment between projected votes $\widehat{\mathbf{u}}_{j|k}^{(i)}$ and the evolving digit capsule representation $\mathbf{S}_j^{(i)}$.
This procedure effectively performs soft clustering in capsule space, reinforcing directionally consistent evidence while attenuating conflicting activations.
Importantly, this agreement mechanism operates on learned embedding geometry rather than assuming explicit geometric part--whole structure within the latent representation of each modality.

The dimensionality of the primary and digit capsules is chosen to ensure that each modality preserves its intrinsic geometric structure. Let $d^{(i)}$ denote the embedding dimension of modality $i$. Projecting $d^{(i)}$ into $N_{\mathrm{pc}}^{(i)}$ capsule vectors of size $d_{\mathrm{pc}}^{(i)}$ ensures that modality-specific variation is not forced into a single subspace, preventing information collapse inherent in early-fusion methods. Moreover, dynamic routing introduces a learnable agreement function that iteratively updates routing coefficients $c_{kj}^{(i)}$ to maximize consistency between lower and higher-level capsules within each modality-specific latent space. This mechanism can be viewed as optimizing a clustering objective in capsule space, where directionally consistent modality-specific evidence is reinforced while noisy or conflicting activations are suppressed. Thus, each architectural element is chosen to preserve modality geometry during routing and provide robust modality-level representations for the subsequent adaptive fusion stage.

\subsubsection{Latent Orientation and Agreement Interpretation}

In classical CapsNet, capsule orientation is often associated with pose parameters in image space.
In our framework, capsules operate entirely in latent embedding space rather than geometric pixel space.
Let $\widehat{\mathbf{u}}_{j|k}^{(i)}$ denote the projected vote from primary capsule $k$ to digit capsule $j$ for modality $i$, and let $\mathbf{S}_j^{(i)}$ denote the resulting digit capsule output.

We interpret the \emph{orientation} of $\mathbf{S}_j^{(i)}$ as the dominant direction of evidence within the learned representation manifold of modality $i$.
Specifically, the direction encodes which latent factors (semantic patterns in text, polarity structure in sentiment, or structural abstractions in image features) are being expressed, while the capsule norm reflects the strength of support for a class hypothesis.

Agreement in routing therefore corresponds to directional consistency between projected votes $\widehat{\mathbf{u}}_{j|k}^{(i)}$ and the evolving digit capsule estimate $\mathbf{S}_j^{(i)}$ within each modality-specific latent space, rather than spatial pose alignment.
This interpretation generalizes capsule orientation from geometric modeling to structured latent evidence alignment within each modality-specific embedding space.

\subsubsection{Confidence Metrics}

The digit capsule outputs are converted into modality-specific confidence scores used as inputs to a downstream gating mechanism. Here, ``confidence'' denotes a summary of capsule magnitude, alignment, or concentration, depending on the modality; it does not denote an empirically calibrated probability of correctness.

For the image modality, the Euclidean norm of each digit capsule is used as a confidence measure, following the capsule-length construction in the original CapsNet formulation. For a nonzero input, the squash function produces a vector with norm strictly below one:
\begin{equation}
\mathrm{squash}(\mathbf{s}) = \frac{\|\mathbf{s}\|^2}{1+\|\mathbf{s}\|^2} \frac{\mathbf{s}}{\|\mathbf{s}\|}, \qquad
z_{\mathrm{img},c} = \bigl\|\mathrm{squash}(\mathbf{S}_{\mathrm{img},c,:})\bigr\|_2,
\end{equation}
With the continuous extension $\mathrm{squash}(\mathbf{0})=\mathbf{0}$, the image score satisfies $z_{\mathrm{img},c}\in[0,1)$.

For the text modality, the objective is to capture semantic alignment between two text-based views (e.g., news categories) of the same underlying region or mortgage context. For nonzero digit capsule outputs, the raw cosine similarity is
\begin{equation}
z_{\mathrm{text}}^{\mathrm{raw}} = \frac{\mathbf{S}_{t_1} \cdot \mathbf{S}_{t_2}}{\|\mathbf{S}_{t_1}\|_2\,\|\mathbf{S}_{t_2}\|_2},
\end{equation}
where $\mathbf{S}_{t_1}$ and $\mathbf{S}_{t_2}$ denote the text digit capsules from two selected news categories. This score lies in $[-1,1]$ and is rescaled to $[0,1]$ via
\begin{equation}
z_{\mathrm{text}} = \frac{1}{2}\left(z_{\mathrm{text}}^{\mathrm{raw}} + 1\right),
\end{equation}
This transformation supplies a bounded alignment score on a common numerical scale with the other modality scores.

For the numeric sentiment modality, a concentration score is derived from the class-wise capsule norms. Let $\ell_c = \|\mathbf{S}_{\mathrm{num},c,:}\|_2$ and
\begin{equation}
p_c = \frac{\exp(\ell_c)}{\sum_{k} \exp(\ell_k)},
\end{equation}
which defines a normalized distribution over classes without establishing probabilistic calibration. The associated entropy,
\begin{equation}
H = - \sum_{c} p_c \log p_c,
\end{equation}
is normalized and converted into a concentration score:
\begin{equation}
z_{\mathrm{num}} = 1 - \frac{H}{\log C},
\end{equation}
where $C>1$ is the number of classes. The score is bounded by zero and one; with finite softmax inputs, it lies in $[0,1)$. A uniform distribution gives zero, whereas a more concentrated distribution gives a higher score. Concentration alone does not establish predictive correctness.

The three scores are bounded within $[0,1]$, although their attainable ranges differ. This common numerical scale is a design convention intended to facilitate their joint use by the gating network. Cosine similarity summarizes alignment for text, capsule length summarizes image evidence, and normalized entropy summarizes concentration for numeric sentiment. Equal numerical values across these scores do not necessarily represent equal reliability or equal probabilities of correct classification.

We use modality-specific summary functions to reflect the different geometric and statistical properties of the encoder outputs. The transformations provide bounded inputs while retaining these distinct interpretations; they do not establish probabilistic calibration or identical score distributions across modalities. The adaptive gate learns how to combine the resulting scores through the final classification objective. No empirical calibration assessment of these intermediate scores or the gating weights is reported in this study.

\subsubsection{Adaptive Weighted Fusion}

We combine the modality-specific confidence scores by rescaling each modality   by a learnable scalar weight $\omega^{(m)}$ for modality $m \in \{\mathrm{text}, \mathrm{img}, \mathrm{num}\}$:
\begin{equation}
\tilde{z}^{(m)} = \omega^{(m)} z^{(m)}.
\end{equation}
The rescaled scores are concatenated into a feature vector
\begin{equation}
\mathbf{f} = [\,\tilde{z}^{(\mathrm{text})} \,\|\, \tilde{z}^{(\mathrm{img})} \,\|\, \tilde{z}^{(\mathrm{num})}\,],
\end{equation}
which is passed through a gating network. The gate first applies a non-linear transformation,
\begin{equation}
\mathbf{h} = \tanh(\mathbf{f}\mathbf{W}_g + \mathbf{b}_g),
\end{equation}
and then converts the result into normalized modality weights using a softmax activation:
\begin{equation}
\boldsymbol{\alpha} = \mathrm{softmax}(\mathbf{h}),
\end{equation}
so that $\alpha_m \ge 0$ and $\sum_m \alpha_m = 1$.

It is important to distinguish the roles of the confidence metrics and the adaptive gating mechanism. The modality-specific confidence metrics summarize evidence within each latent representation, while the adaptive gate learns, through end-to-end optimization, how much each already-computed confidence signal should contribute to the final prediction. Consequently, the gating mechanism does not determine how confidence is measured; instead, it adaptively combines normalized modality-specific summary scores from heterogeneous modalities into a unified decision.

The adaptive gating mechanism represents the first stage in the framework where modality-level integration is explicitly introduced. While the routing procedure operates independently within each modality, the gating network jointly processes modality-level confidence signals, enabling instance-specific weighting and integration of information from multiple modalities. Therefore, consistency of modality-level evidence in this framework is realized through the integration of modality-level evidence summarized by normalized confidence scores rather than direct capsule-level alignment.

The gating network is trained jointly with the classification head; therefore, the modality weights $\boldsymbol{\alpha}$ are learned directly through backpropagation to minimize the final loss. The intermediate $\tanh(\cdot)$ transformation provides non-linear modeling capacity, while the subsequent softmax ensures that the resulting weights form a normalized distribution over modalities. This combination yields an adaptive weighting mechanism, where $\alpha_m$ represents the relative weight assigned to modality $m$ for an instance. These softmax outputs are normalized gating weights, not calibrated probabilities that a modality is correct.

The adaptively weighted representation can then be combined with the output of the digit capsule or passed to a final classification head (described in the following subsection) to produce the mortgage-default risk prediction. This fusion mechanism preserves the modality-specific structure and operates on normalized and comparable confidence signals. Provides a transparent means of assessing the relative contribution of each data source.


The final prediction is obtained by passing the gated representation through a linear layer that outputs the estimated default probability $\hat{y}$. The entire model is trained end-to-end using binary cross-entropy,
\[
\mathcal{L} = -\left(y\log \hat{y} + (1-y)\log(1-\hat{y})\right),
\]
which ensures that the fusion mechanism and routing parameters are optimized jointly for the mortgage-risk classification objective.

\subsection{Implementation}
\label{sec:implementation}

We present the details of the implementation of the two-phase framework shown in Figure~\ref{fig:framework} that features text preprocessing, sentiment extraction, category selection, and selection of unimodal models prior to multimodal fusion. In the Unified Analysis phase, we process news articles to extract sentiment-based and semantic signals and compute sentiment scores at the document level using the VADER model (Valence Aware Dictionary and sEntiment Reasoner)~\cite{isnan2023sentiment}. VADER is a lexicon and rule-based tool tailored to short social media and news texts. VADER produces polarity scores (positive, neutral, negative) and a composite index that summarizes the overall sentiment intensity. For each keyword-defined news category, we average sentiment scores across all documents in that category, yielding an aggregate sentiment profile. These aggregated scores form a numeric modality that complements text embeddings by capturing an emotional tone that may not be fully represented in the topic or semantic structure.
We embed documents using the Universal Sentence Encoder (USE)~\cite{li2022brief} 
 to identify which text categories contribute the most distinct information. USE is a deep neural network model trained over a variety of corpora and tasks to generate fixed-length, semantically meaningful vector representations of sentences. We obtain category-level representations by aggregating document embeddings within each category. We compute the cosine similarity between these category vectors, as this similarity metric quantifies the extent of overlap or independence between categories in the USE-induced semantic space. We retain two categories that exhibit low mutual similarity, providing semantically complementary text channels. This selection strategy reduces computational overhead while preserving diverse perspectives, as illustrated in Figure~\ref{fig:framework}.

We carry out an independent model-selection step for each modality within the same Unified Analysis phase. We train multiple candidate architectures that are evaluated separately for text, numeric sentiment, and LiDAR-based image data using a held-out validation set. For each modality,  we retain the best-performing model according to a predefined performance metric (e.g., AUC)  as the modality-specific encoder. These encoders define the interface between raw inputs and the multimodal fusion stage.

In the Multimodal Analysis phase, we integrate the selected encoders into the FusionCapsNet joint architecture as described in Section~\ref{sec:fusion}. We first transform the text, sentiment, and image inputs into latent representations by their respective encoders and subsequently pass them to the modality-specific capsule-routing stage, followed by the adaptive gating mechanism. This implementation strategy exploits the strengths of unimodal optimization while enabling the fused model to adaptively integrate modality-level evidence summarized by normalized confidence scores in a controlled and interpretable way.

\subsection{Data}

Our study utilises residential mortgage data from the Netherlands, obtained through the European Data Warehouse (EUDW) \cite{EUDW2022}. We focus on three mortgage portfolios that have been selected based on the availability of complete and consistent records, covering the period from 2013 to 2022.

During preprocessing, we removed records with missing geographic identifiers, inconsistent payment histories, and duplicate entries to ensure label reliability and data integrity. The experimental cohort contains approximately 14,000 mortgage records, of which 15\% are classified as high-risk and 85\% as low-risk. The high-risk class includes both Arrears and Default or Foreclosure.

Although our multimodal fusion framework is designed to operate using cost-free and publicly accessible unstructured data sources—such as financial news text and LiDAR-based spatial imagery—the present empirical study combines these inputs with mortgage-level records obtained from the European Data Warehouse (EUDW) for supervised evaluation. In this context, “cost-free” refers to the absence of borrower-level data purchase, per-record pricing, or proprietary licensing agreements, rather than the complete absence of any access or administrative cost.

Importantly, the EUDW data are used solely to provide ground-truth labels for benchmarking predictive performance and are not required as input to the proposed framework. Compared with commercial credit-bureau data, whose access, cost, and legal requirements vary substantially across jurisdictions, the EUDW represents a standardized and relatively low-barrier source that enables empirical validation without reliance on proprietary scorecards. Accordingly, while restricted loan-level data are used for evaluation in this study, the framework itself does not depend on privileged or institution-specific financial inputs.

The target variable in our analysis is \emph{loan performance}, classified into five distinct statuses: ``Performing'', ``Arrears'', ``Default or Foreclosure'', ``Redeemed'', and ``Repurchased by Seller''. The primary objective is to predict \emph{default (or high-risk) rate}, a key performance indicator for mortgage lenders. To simplify the prediction task, we aggregate the ``Arrears'' and ``Default or Foreclosure'' categories into a single \emph{high-risk} group. This grouping is justified, as both statuses represent financial distress in which borrowers are unable to make timely mortgage payments. The EUDW indicates that most of the loans in their system have a lower default rate than the system, so detecting risky operations to identify the overall risk of the segment, area, or customer segment.

In contrast, the categories ``Performing'', ``Redeemed'', and ``Repurchased by Seller'' are considered \emph{low-risk}, indicating regular repayment, full loan redemption, or repurchase by the originating seller. Formally, we define:
\begin{itemize}
    \item \textbf{High-risk mortgages:} ``Arrears'' or ``Default or Foreclosure''
    \item \textbf{Low-risk mortgages:} ``Performing'', ``Redeemed'', or ``Repurchased by Seller''
\end{itemize}

\begin{figure}[htbp!]
\centering
{\includegraphics[width=2in]{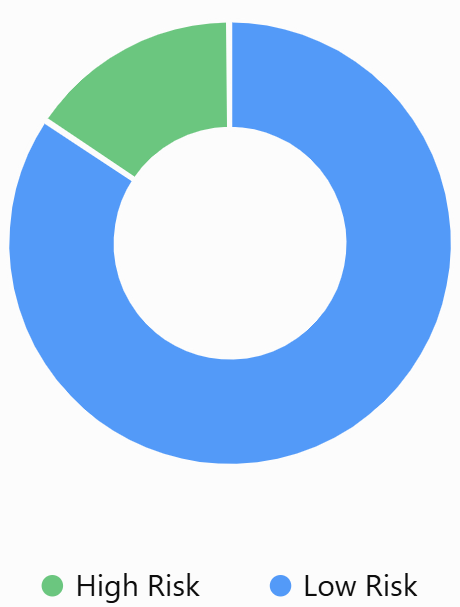}}
\caption{Class distribution of the experimental mortgage cohort (approximately 14,000 records). The high-risk group (arrears or default/foreclosure) constitutes 15\% of observations, and the low-risk group constitutes 85\%.}

\label{fig:distribution}
\end{figure}

The distribution of these two classes is illustrated in Figure~\ref{fig:distribution}. 

To eliminate the confounding effect of time on loan performance, we construct the dataset such that all mortgages are classified as ``Performing'' at the beginning of the observation period. We then follow their status three years later to determine whether they fall into the \emph{low-risk} group (``Performing'', ``Redeemed'', or ``Repurchased by Seller'') or the \emph{high-risk} group (``Arrears'' or ``Default or Foreclosure'').

\subsubsection{Text Data} 
\begin{figure}[htbp!]
\centering
\includegraphics[width=3.3in]{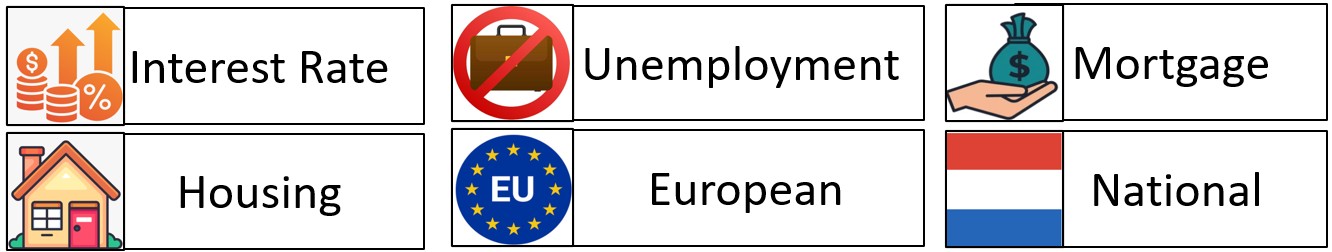}
\caption{Selected keyword categories for text data extraction, representing economic, housing, and policy factors potentially influencing mortgage risk.}
\label{fig:category}

\end{figure}

To explore the relationship between news content and mortgage risk, we extracted targeted news articles from the Seeking Alpha platform \ \cite{seekingalpha}, a well known source for providing global economic and financial news. 
The extraction was based on a set of keywords relevant to factors that could influence lenders' risk assessment when approving mortgages.  Figure~\ref{fig:category} illustrates the keywords that are selected under the assumption that they have a significant impact on lending decisions on high-risk mortgages.

 We constructed a set of keywords to identify the most relevant news for our analysis, which reflect economic and policy factors closely related to mortgage risk. Among these, interest rates are particularly pivotal: Lower rates often encourage lenders to approve higher risk mortgages \cite{gill2023health, magri2011rise}, whereas higher rates usually make lenders more cautious. The conditions of the housing market also play a central role: periods of rising prices and strong demand may reduce the perceived collateral risk, while declining markets or oversupply increase the probability of default \cite{hodge2017assessment}. Unemployment rates further shape lender behavior, as high unemployment signals economic uncertainty and vulnerability of the borrower, prompting more conservative lending \cite{gabriel1991credit, gyourko2014reconciling}. In addition, European-level housing and mortgage regulations influence risk taking by imposing requirements on income verification, loan-to-value ratios, and debt-to-income limits \cite{mak2015responsible}. Finally, we include the keyword ``Netherlands'' to ensure that country-specific economic and policy news is captured within the Dutch context.


We aligned news articles and their associated sentiment scores  with the 
mortgage origination timeline to ensure temporal validity and avoid leakage of forward-looking information. Specifically, we included only news published within a three- to six-month window prior to the mortgage observation date. This was done to ensure that all text-based features reflect information that would have been available to lenders at the time of their risk assessment.

\subsubsection{Image Modality} 

\begin{figure}[htbp!]
\centering
\includegraphics[width=3.5in]{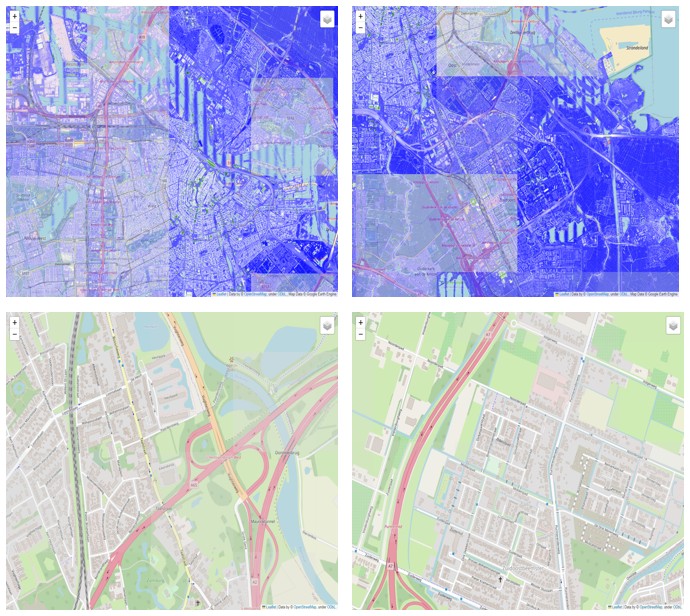}
\caption{Sample LiDAR maps illustrating elevation and spatial features, including green space, building density, water bodies, and transport infrastructure.}
\label{lidar}
\end{figure}

We incorporated LiDAR imagery from the Earth Engine Data Catalog \cite{AHN2_2022} as an additional data source in our study. 
Specifically, we used the \textit{Current Height Model of the Netherlands} (AHN)\footnote{In Dutch: \textit{Actueel Hoogtebestand Nederland}}, which provides a high-resolution Digital Elevation Model (DEM) \cite{gupta2018digital} derived from LiDAR data. The AHN DEM offers detailed topographic information for the Netherlands, capturing both ground-level features and above-ground structures such as buildings, bridges, and vegetation (Figure~\ref{lidar}). We created a dataset by converting LiDAR point clouds into a 0.5\,m grid using a squared inverse distance weighting method, allowing for a precise representation of the elevation of the terrain.  
To integrate this data with our mortgage dataset, we employed a geospatial matching process based on the postal code of each loan, using latitude and longitude coordinates to retrieve the corresponding DEM segment. This allowed us to extract spatial features such as building density, proximity to water bodies, vegetation coverage, and transportation infrastructure, all of which may correlate with the risk of default on the mortgage.

Although news articles and LiDAR maps provide unstructured input, the sentiment scores extracted from the text and the EUDW loan-performance variables are structured numerical features. These structured elements complement the unstructured sources and are expected in a supervised prediction setting.

\subsection{Experimental Design}

\subsubsection{Text Category Selection and Sentiment Feature Extraction}

\begin{figure}[htbp!]
\centering
{\includegraphics[width=3.5in]{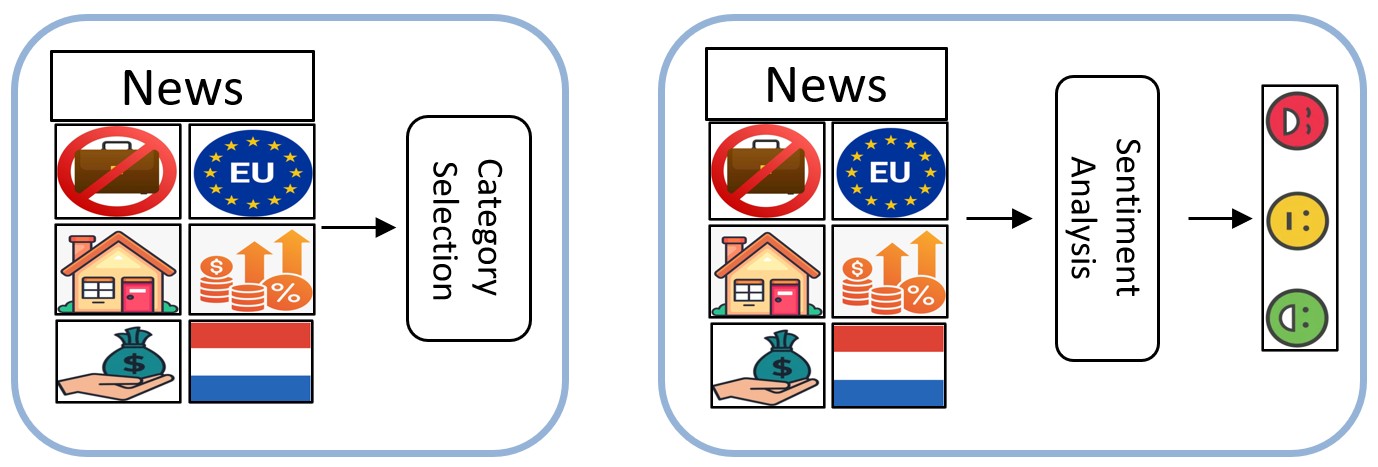}}
\caption{Illustration of the text processing pipeline in the Unified Analysis phase. 
The left panel depicts the \textit{Category Selection} step, where semantically independent news categories are identified to reduce redundancy and noise. 
The right panel shows the \textit{Sentiment Analysis} step, in which each selected category is assigned a sentiment score to capture the emotional tone of the news content.}
\label{fig:textanalysis}
\end{figure}

We address the problem that certain news categories can carry overlapping or highly correlated information in the initial stage of the Unified Analysis~ (Figure \ref{fig:textanalysis}), which can introduce redundancy and noise into the modeling process. To mitigate this, we aim to select semantically independent categories, thereby preserving diverse and non-redundant thematic perspectives. We employ cosine similarity to quantify the semantic independence between categories by comparing their document embeddings. This metric is particularly suitable as it measures the angular distance between vectors, making it robust to variations in text length and scale. Based on the similarity matrix, we identify ``Housing'' (most similar to others) and ``Netherlands'' (most distinct) to balance semantic richness and diversity. Following category selection, we extract sentiment scores from the news text using the VADER model. These sentiment features, in addition to the original text, are incorporated into the multimodal phase to enhance the representation of textual information, enabling the model to capture both the semantic content and the emotional tone of the news articles.

\subsubsection{Model Selection}

\begin{figure}[htbp!]
\centering
{\includegraphics[width=3.5in]{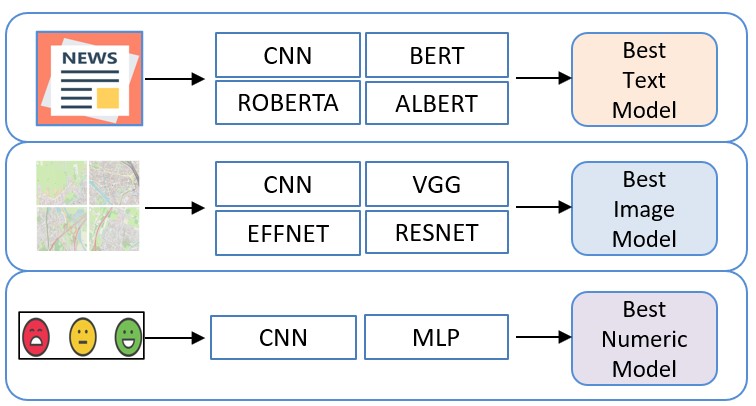}}
\caption{The model selection process in the unified analysis phase, where we train multiple models for each data channel (text, image, and numerical features). The top-performing model from each channel is selected for the subsequent multimodal analysis phase.}
\label{fig:model selection}
\end{figure}

In the unified analysis phase, we evaluate multiple well-known models for each data channel to identify the most suitable one for the multimodal analysis stage~ (Figure \ref{fig:model selection}). For each channel, several candidate models are trained and evaluated on their performance. The model with the highest performance for each channel is then selected to be integrated into the subsequent multimodal fusion process.

\subsubsection{Multimodal Analysis}

In the second phase, we construct the multimodal framework using the results obtained from the unified analysis. As illustrated in Figure~\ref{fig:framework}, Phase~2, we process each input channel through its channel-specific channel encoder that performs best: the text encoder for textual inputs, the numeric encoder for sentiment score features, and the image encoder for visual data. The output of these encoders is then integrated in the fusion stage.  
We use our FusionCapsNet framework, which enables structured agreement modeling within each modality-specific embedding space. This design preserves modality-specific representational diversity while supporting adaptive modality-level integration through the aggregation of modality-level evidence summarized by normalized confidence scores, thereby improving interpretability in the prediction process. The fused representation is subsequently used to generate the final risk prediction.

\section{Results}
\label{sec:result}
\subsection{Sentiment Analysis}

As discussed above, in addition to leveraging the semantic information of the text channel, we extract sentiment scores as a separate numeric modality. 
This approach allows us to capture the emotional tone embedded in the news content, providing complementary information that may not be fully reflected in semantic embeddings alone. 
We can incorporate emotional signals, such as optimism, by quantifying sentiment, uncertainty, or negativity, into our predictive model.  This enables it to detect risk patterns that arise not only from the topics being discussed but also from the tone in which they are presented.

It is worth noting that, while the number of articles naturally varies across categories, this does not create bias in our approach.
Each text contributes a sentiment score for each category, so all categories are uniformly represented in the sentiment feature space.

Table~\ref{average} presents the average sentiment scores for the news documents, calculated using VADER as described in Section~\ref{sec:methodology}.
These scores reveal the prevailing sentiment tendencies within each keyword category. 
Our objective in extracting sentiment scores is not to perform sentiment classification, but rather to construct a numeric channel that captures the emotional tone embedded in the news text. 
In this context, the average sentiment values across categories are not meant as standalone predictive findings, but as descriptive indicators that help illustrate relative differences between categories. 
This comparison provides intuition about which topics are framed more negatively or positively in the news, while the actual role of sentiment in our framework is as a continuous input feature that complements other modalities.
The ``Netherlands'' category has an average sentiment score of 0.101, indicating a slightly positive tone, suggesting that country-specific news often expresses marginally more positivity than negativity. 
In contrast, the ``Interest rate'' category has a moderately positive sentiment score of 0.287, implying that related articles often emphasize favorable economic conditions, borrowing advantages, and potential investment benefits.  
The ``Mortgage'' category records an average sentiment score of -0.055, reflecting a slightly negative tone that may arise from discussions about the challenges, risks, or difficulties faced by the borrower today in the country. 
Similarly, ``Housing market'' has a sentiment score of -0.038, suggesting a mild negativity possibly related to affordability problems, market volatility, or supply constraints.  
The ``Unemployment'' category stands out with a strongly negative score of -0.391, reflecting consistent coverage of job losses, economic instability, and associated social impacts \cite{shayaa2017social}.  
Finally, ``European policy'' shows a mildly negative sentiment score of -0.060, indicating mixed or critical views on regulatory and policy developments. The remaining categories tend to hover around neutrality, reflecting a balanced mix of positive and negative reporting.  

\begin{table}[!ht]
\begin{center}
\caption{The average sentiment scores calculated using VADER model for news documents across different keyword categories reveal insights into the prevailing sentiments.
}
\label{average}
\begin{tabular}{cc}
\toprule
Text Category & Average Score \\\midrule\midrule
Netherlands &0.101\\
Interest rate  &0.287\\
Mortgage  &-0.055\\
Housing market  &-0.038\\
Unemployment &-0.391\\
European policy  & -0.060\\\bottomrule
\end{tabular}
\end{center}
\end{table}

\subsection{Text Category Selection}

As outlined in the dataset section, our sentiment analysis model initially considers six diverse categories of text to ensure a comprehensive use of the available information. However, we narrow this to two specific categories that yield the best predictive performance to strike a balance between maintaining strong performance and adhering to computational resource constraints while still enabling informed decision-making based on the available text data.  
As described in Section~\ref{sec:framework}, we identify the two categories by computing cosine similarities among all candidate text representations and selecting those that are the most semantically independent.
The results in Table~\ref{tab:cosine_similarity} show that the documents containing the keywords \textit{Mortgage} and \textit{Housing} exhibit the highest similarity score (0.544), suggesting substantial semantic overlap. In contrast, documents associated with the keyword \textit{Netherlands} display the lowest similarity to all other categories, particularly with \textit{Housing} and \textit{Mortgage}.  
Based on these findings, we selected \textit{Mortgage} and \textit{Netherlands} as the two most independent categories for further analysis. The \textit{Mortgage} category captures direct and domain-specific signals related to housing finance risk, while the \textit{Netherlands} category introduces geographically specific macroeconomic and policy-related information. This combination enables our model to benefit from both highly targeted mortgage-related content and broader, independent country-level factors, thereby maximizing informational diversity while avoiding redundancy across text inputs.

\begin{table*}[!ht]
\centering
\caption{Cosine similarity scores between different text documents.}
\label{tab:cosine_similarity}
\resizebox{\textwidth}{!}{%
\begin{tabular}{@{}lcccccc@{}}
\toprule
& \textbf{Unemployment} & \textbf{European} & \textbf{Housing} & \textbf{InterestRate} & \textbf{Mortgage} & \textbf{Netherlands} \\
\midrule
\textbf{Unemployment} & 1.000 & 0.270 & 0.206 & 0.369 & 0.248 & 0.069 \\
\textbf{European} & 0.270 & 1.000 & 0.268 & 0.269 & 0.374 & 0.070 \\
\textbf{Housing} & 0.206 & 0.268 & 1.000 & 0.294 & 0.544 & -0.003 \\
\textbf{InterestRate} & 0.369 & 0.269 & 0.294 & 1.000 & 0.345 & -0.060 \\
\textbf{Mortgage} & 0.248 & 0.374 & 0.544 & 0.345 & 1.000 & 0.041 \\
\textbf{Netherlands} & 0.069 & 0.070 & -0.003 & -0.060 & 0.041 & 1.000 \\
\bottomrule
\end{tabular}%
}
\end{table*}

\subsection{Image vs numeric data streams}

In our multimodal fusion framework, we incorporate spatial (map) data that has been associated with mortgage data using postal codes as a supplementary channel. However, before using the image data in our multimodal model, we prioritize ensuring that the performance of the visual data significantly exceeds the performance of the postal code data. Table~\ref{auc} presents the comparison of spatial models (LiDAR map)  and numeric data (Postal Code) using the Area Under the Curve (AUC). We find that the spatial model outperforms the numeric model, which indicates that the spatial data contains more information and contributes significantly to the model's predictive power. A reason for this could be that postal codes can be broad and cover a wide area, potentially missing out on specific local factors that can influence mortgage-related patterns and outcomes.

\begin{table}[!ht]
\begin{center}
\caption{Comparison between spatial (LiDAR map)  and numeric data (Postal Code) models using AUC.}
\label{auc}
\begin{tabular}{cc}
\toprule
Channel& AUC \\\midrule\midrule
Numeric Postal Code &0.516$\pm$0.013\\
LiDAR Map & 0.607$\pm$0.039\\\bottomrule
\end{tabular}
\end{center}
\end{table}

\subsection{Unimodal selection }

In this section, our approach involves training individual channels (modalities) independently before merging the best models from each channel into a multimodal framework. We evaluate the performance of each channel using three criteria. The first two are the well-known Area Under the ROC curve (AUC) and the F1 score. 
 AUC is particularly useful in imbalanced settings because it evaluates the ability of a model to discriminate between classes on all thresholds, rather than being influenced by the majority class. The F1 score complements this by providing a single measure that balances precision and recall, which is also important when the positive class is underrepresented.

In addition, we used the partial AUC (pAUC) to assess the performance of the model at specific thresholds. These thresholds are defined in terms of the false positive rate (FPR), and we restrict attention to the low-FPR region (for example, FPR $\leq$ 0.10), which is most relevant for the prediction of mortgage default. High FPR quickly become economically infeasible, hence no rational player would offer these cutoffs to the public. In this setting, even a small number of false positives can have material implications for lenders, so focusing on the early part of the ROC curve provides a more realistic evaluation of model profitability. By reporting both the full AUC and the pAUC, we capture the overall ranking ability and also highlight the performance in the operating region of the greatest practical importance.

Table~\ref{unimodal} summarizes the performance of the unimodal model in terms of AUC, pAUC, and F1 scores for three channels: Text, Image, and Numeric.  In the case of the Text channel, we compare CNN, BERT, RoBERTa, and ALBERT. We find that BERT achieves the highest performance, outperforming the other models. The relatively lower performance of RoBERTa and ALBERT compared to BERT may be attributed to differences in their pre-training strategies. Although RoBERTa and ALBERT are variants of BERT designed to improve performance in various NLP tasks, these modifications may not be optimally aligned with the requirements of predicting mortgage defaults from news text. BERT's pre-training and fine-tuning process appears to align more closely with the linguistic and contextual characteristics of our task, giving it a performance advantage.

In the case of the Image channel, we compare CNN, VGG, EfficientNet, and ResNet. We find that VGG achieves the best performance, surpassing the other models in all evaluation metrics.
This stronger performance can be attributed to VGG's deeper architecture and uniform convolutional design, which capture fine-grained spatial features such as edges, textures, and structural patterns that are particularly relevant in LiDAR-based images. In contrast, EfficientNet and ResNet are optimized for more complex or large-scale visual tasks. Furthermore, CNN lacks the depth to fully capture high-level spatial cues, making VGG better aligned with the requirements of our mortgage risk prediction task.

Finally, in the numerical channel, which includes sentiment scores, we evaluate CNN and MLP. The MLP model consistently outperforms CNN, indicating its superior ability to model numeric features in this context. We included CNN in this comparison for consistency across modalities, as our methodology involves systematically evaluating alternative models within each channel before selecting the best-performing one.

Overall, these results indicate that BERT excels in capturing the semantic and contextual nuances of textual data, VGG is highly effective for visual feature extraction, and MLP is well-suited for processing structured numeric information. This motivates our multimodal framework to combine these top-performing models, leveraging their complementary strengths to improve predictive accuracy.

\begin{table*}[!ht]
\centering
\caption{Comparison of Unimodal Model Performance for Predicting Mortgage Defaults}
\label{unimodal}
\begin{tabular}{l l c c c}
\toprule
Channel & Model & AUC & PAUC & F1 \\
\midrule\midrule
\multirow{4}{*}{Text} 
& CNN     & 0.595$\pm$0.038 & 0.641$\pm$0.018 & 0.510$\pm$0.025 \\ 
& BERT    & 0.604$\pm$0.037 & 0.664$\pm$0.007 & 0.516$\pm$0.012 \\
& RoBERTa & 0.534$\pm$0.034 & 0.656$\pm$0.015 & 0.471$\pm$0.004 \\
& ALBERT  & 0.524$\pm$0.030 & 0.655$\pm$0.016 & 0.461$\pm$0.004 \\
\midrule
\multirow{4}{*}{Image} 
& CNN  & 0.535$\pm$0.039 & 0.632$\pm$0.018 & 0.460$\pm$0.009 \\ 
& VGG     & 0.607$\pm$0.034 & 0.673$\pm$0.014 & 0.496$\pm$0.020 \\
& EffNet  & 0.566$\pm$0.032 & 0.658$\pm$0.014 & 0.462$\pm$0.004 \\
& ResNet  & 0.578$\pm$0.033 & 0.666$\pm$0.014 & 0.469$\pm$0.003 \\
\midrule
\multirow{2}{*}{Numeric} 
& CNN     & 0.521$\pm$0.031 & 0.625$\pm$0.015 & 0.461$\pm$0.030 \\ 
& MLP     & 0.571$\pm$0.033 & 0.637$\pm$0.015 & 0.481$\pm$0.030 \\
\bottomrule
\end{tabular}
\end{table*}

\subsection{Multimodal Analysis}
\label{sec:fusion}
\begin{figure*}[htbp!]
\centering
{\includegraphics[width=6.5in]{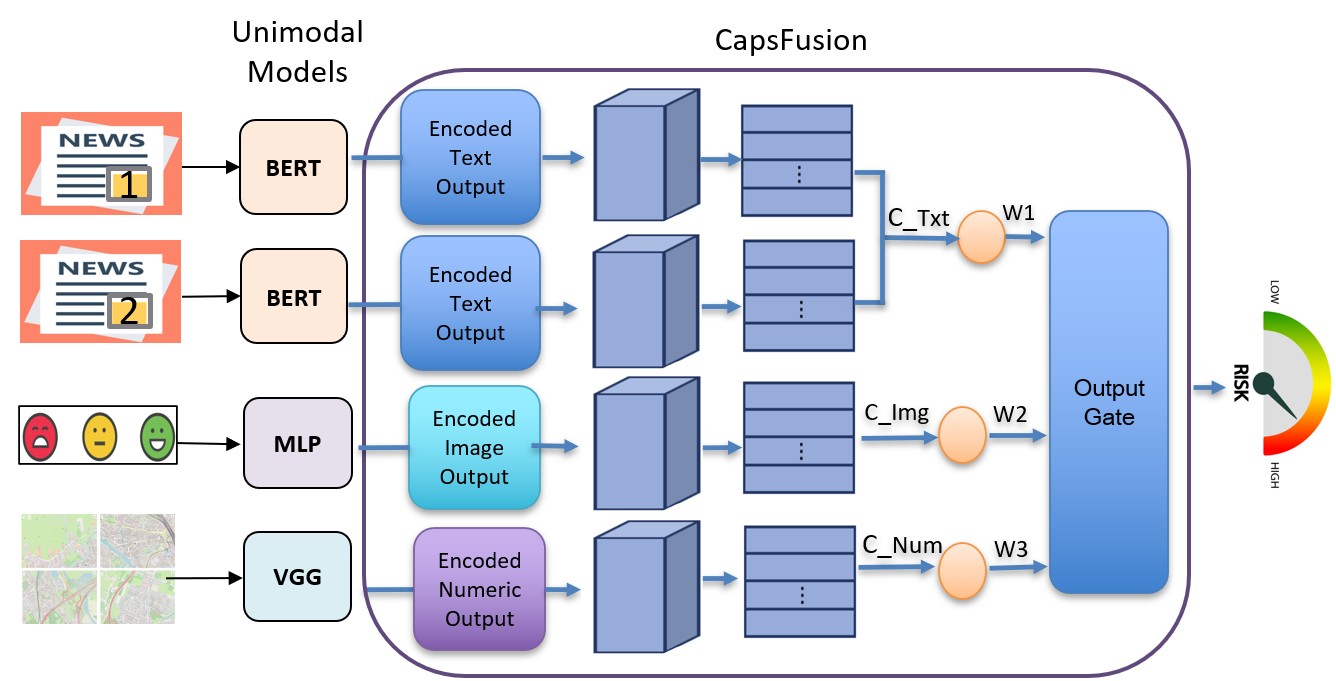}}
    \caption{FusionCapsNet framework for multimodal mortgage risk prediction featuring unimodal models (BERT for text, VGG for images, and MLP for sentiment scores) to generate encoded representations. We expand these outputs into primary capsules, processed through digit capsules with routing-by-agreement independently within each modality, and evaluated using modality-specific confidence metrics. The resulting scores are integrated across modalities through adaptive gating to produce the final risk prediction.}

\label{fig:final}
\end{figure*}

We design our multimodal framework by selecting the best-performing model for each channel based on the Unimodal results. As shown in the previous section, we used BERT for the text modality, VGG for the image modality, and MLP for the numeric modality, which consists of sentiment scores extracted from the text data. These modality-specific models generated high-quality feature representations, which are integrated in the fusion stage using our proposed \textit{Caps Fusion} approach to predict the risk of default on the mortgage shown in Figure~\ref{fig:final}.

 We compare our FusionCapsNet's performance with three widely used fusion strategies identified in the literature review: \textit{Addition}, \textit{Concatenation}, and \textit{Cross-Attention}.
The \textit{Addition} fusion method combines modality-specific feature vectors by adding elements in a way that effectively averages the contributions of each modality. The \textit{Concatenation} method merges the feature vectors by stacking them into a single longer vector, allowing subsequent layers to learn weighted combinations of all features together.

Our multimodal structure is in line with modern developments. The \textit{Cross-Attention} fusion model implements a transformer-based attention mechanism that learns inter-modal dependencies by allowing each modality to attend to the most informative features from the others, representing a state-of-the-art baseline for multimodal alignment.

To avoid information leakage, we partitioned the data before preprocessing, feature extraction, and model training. We partitioned the dataset  into training, validation, and test sets using an 80/10/10 split at the mortgage-record level, ensuring that each mortgage and its associated text, sentiment, and LiDAR features appear in only one split.

To prevent temporal leakage, the dataset was split prior to any preprocessing or feature extraction. 
All features were constructed using only information available before mortgage origination, including news published within a 3–6 month window prior to each loan. 
No post-origination variables or future macroeconomic indicators were used in model training.

Table~\ref{performance} compares the performance of four fusion strategies: FusionCapsNet, Cross-Attention, Addition, and Concatenation -- on multimodal mortgage default prediction using AUC, pAUC, and F1 scores. We find that FusionCapsNet achieves the highest overall performance, indicating its strong ability to capture complementary information across modalities and maintain discriminative power.
We observed that the Cross-Attention model achieved higher training AUC but exhibited greater variance across runs, suggesting sensitivity to sample size and regularization strength. 

The Cross-Attention model achieved higher training performance and uses projection into a shared latent space (dimension = 128) together with multiple attention parameters. However, it also showed higher variance across runs, indicating sensitivity to training conditions and random initialization. One possible explanation is sensitivity to the amount and alignment of training information relative to model capacity. The inputs comprise two selected text views, LiDAR imagery, and numeric sentiment, grouped into three modality types. The present experiments do not isolate data scale as the cause of the performance gap. The model was trained using Adam (learning rate = $3 \times 10^{-4}$), dropout = 0.1, gradient clipping (clipnorm = 1.0), batch normalization, and early stopping based on validation AUC.

Despite its higher capacity, the Cross-Attention model did not outperform simpler fusion methods such as concatenation, likely due to limited data scale and relatively weak alignment between modalities, under which attention mechanisms may struggle to learn robust cross-modal interactions. 
Importantly, no artificial constraints were imposed on the baseline, and all models were evaluated under the same preprocessing pipeline, data splits, optimizer settings, and evaluation framework to ensure fair comparison. To support transparency and reproducibility, the full implementation code for FusionCapsNet and all baseline models is provided in the GitHub repository referenced in the Code Availability section of the manuscript.

\begin{figure}[htbp]
\centering
\includegraphics[width=0.88\linewidth]{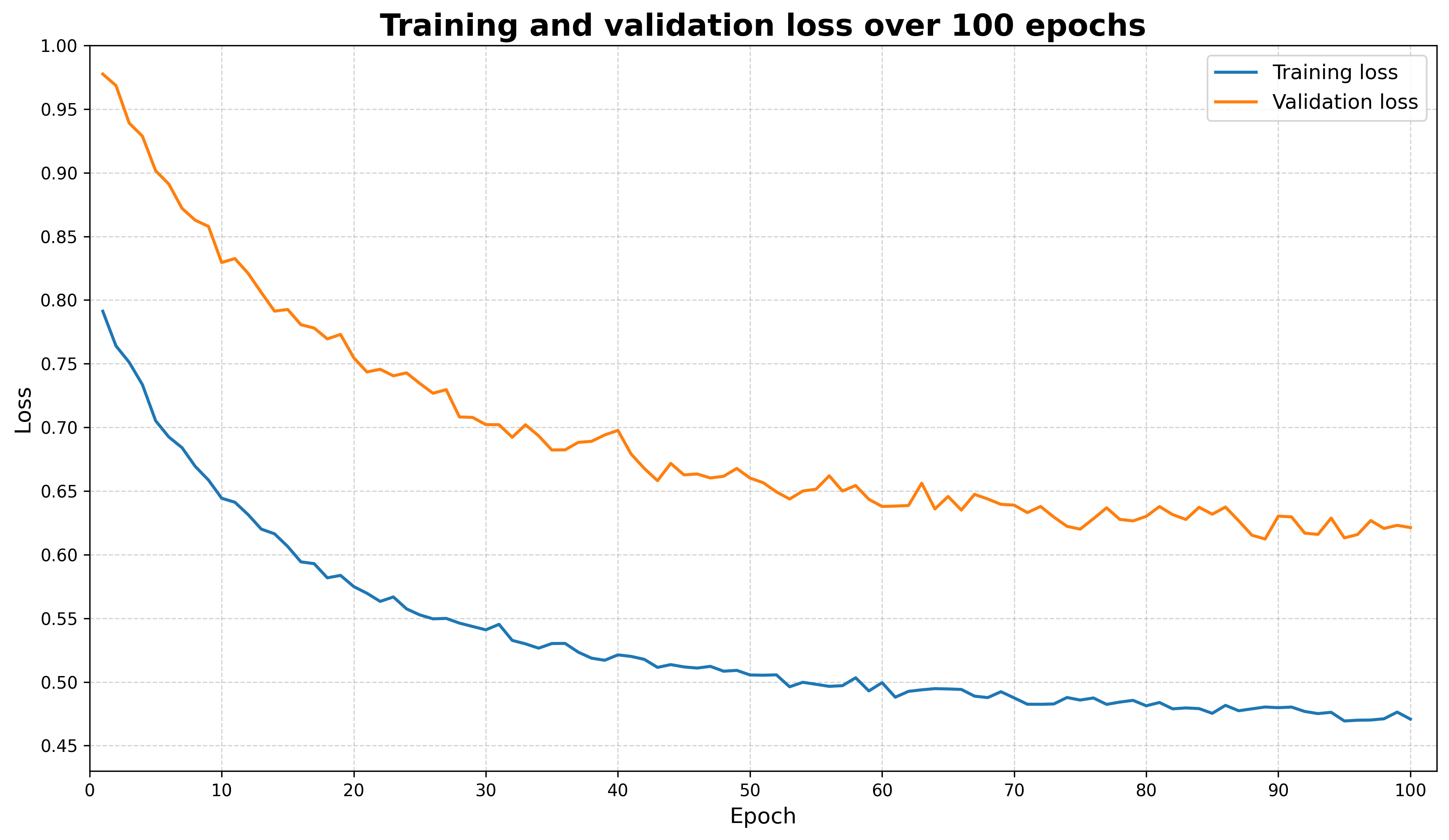 }
\caption{Training and validation loss trajectories of the Cross-Attention baseline across training epochs.}
\label{fig:cross_attention_convergence}
\end{figure}

The comparatively lower performance of the Cross-Attention baseline should not be interpreted as evidence that cross-attention is generally ineffective. As shown in Figure~\ref{fig:cross_attention_convergence}, both the training and validation losses converge stably, indicating that the optimization procedure was successful and that the model did not suffer from training instability or divergence. Instead, we attribute the reduced predictive performance to the characteristics of the present dataset. Cross-attention is typically most effective when strong fine-grained correspondences exist across modalities, such as word–region alignments in vision-language applications. In our application, however, financial news, LiDAR-derived spatial features, and aggregated sentiment represent heterogeneous information sources with relatively weak instance-level correspondence. Consequently, early cross-modal attention may introduce noisy interactions that limit generalization, whereas preserving modality-specific representations before adaptive fusion enables more robust integration of complementary information.

We report the mean $\pm$ standard deviation in 30 repeated independent experimental runs, obtaining the respective metrics. The small variance observed in AUC, pAUC, and F1 indicates that gains are reproducible between random initializations, providing partial robustness evidence even in the absence of formal statistical significance tests.

To further contextualize the performance of FusionCapsNet, based on public industry reports, we note that commercial and regulatory mortgage-risk scorecards typically achieve AUC values in the 0.60–0.70 range. Since FusionCapsNet consistently achieves AUC scores within this same band (0.66–0.69), the results provide quantitative evidence that the model approximates the predictive performance of standard scorecard benchmarks, despite relying heavily on unstructured data sources.


\begin{table}[!ht]
\centering
\vspace{-0.2cm}
\caption{Performance and approximate training time of different fusion strategies for multi-modal mortgage default prediction. We measured the model training time using a T4 GPU}
\label{performance}
\adjustbox{max width=0.95\textwidth}{
\begin{tabular}{lcccc}
\toprule
Fusion Strategy & AUC & pAUC & F1 &  Time (hours)$^{\dagger}$ \\
\midrule\midrule
FusionCapsNet   & 0.692$\pm$0.021 & 0.696$\pm$0.014 & 0.552$\pm$0.018 & $\sim$2.2 \\
Cross-Attention & 0.607$\pm$0.030 & 0.649$\pm$0.011 & 0.493$\pm$0.025 & $\sim$3.5 \\
Addition        & 0.622$\pm$0.013 & 0.664$\pm$0.017 & 0.511$\pm$0.002 & $\sim$1.5 \\
Concatenation   & 0.668$\pm$0.021 & 0.682$\pm$0.019 & 0.547$\pm$0.012 & $\sim$1.5 \\
\bottomrule
\end{tabular}}
\end{table}

We performed paired  statistical significance tests on 30 repeated runs using identical data splits to assess whether performance differences are meaningful.  A paired t-test confirms that FusionCapsNet significantly outperforms all baseline fusion strategies on AUC, pAUC, and F1 (all $p < 0.05$). 
Thus, the observed performance gains are not attributable to random variation.

As an ablation study, we compare the strongest unimodal models for each channel 
(BERT for text, VGG for LiDAR images, and MLP for numeric sentiment) against the 
proposed multimodal FusionCapsNet (Tables~\ref{unimodal} and \ref{performance}). 
Individually, the text (AUC = 0.604), image (AUC = 0.607), and sentiment (AUC = 0.571) channels achieve moderate performance, confirming that each modality encodes non-redundant information related to mortgage risk. When fused, FusionCapsNet increases AUC to 0.692, demonstrating that the modalities are complementary and that multimodal integration substantially improves predictive usefulness. This ablation evidence supports the effectiveness of the fusion mechanism in adaptively integrating modality-level evidence summarized by normalized confidence scores, thereby exploiting complementary information from multiple modalities that unimodal models cannot capture in isolation.

\paragraph{Ablation Study: Isolating the Routing Mechanism}
To verify that the performance gain arises from agreement-based routing rather than increased parameter capacity, we conduct controlled ablations while keeping unimodal encoders (BERT, VGG, and MLP), capsule dimensionality, and training protocol fixed.
When dynamic routing is replaced with fixed average pooling across primary capsule votes (AvgPoolCaps), the AUC decreases to 0.663. 
When the capsule structure is removed entirely and replaced with a parameter-matched dense fusion layer (DenseFusion), performance further declines to an AUC of 0.648. 
These results indicate that the improvement of FusionCapsNet is not solely attributable to increased parameter capacity. Rather, the iterative routing-by-agreement mechanism contributes measurably to discriminative performance by reinforcing directionally consistent latent evidence within each modality-specific embedding space, which is subsequently integrated through the adaptive gating mechanism.

\subsection{Out-of-Time (OOT) Validation}

To further assess temporal robustness, we conducted an out-of-time (OOT) experiment using a strict chronological split, training the model on mortgages originated between 2013–2020 and testing on 2020–2022. 
Under this setting, FusionCapsNet achieves an AUC of 0.674. 
Although slightly lower than the 0.692 AUC observed under random splitting, the performance remains stable, indicating that the model captures structurally meaningful signals rather than benefiting from temporal leakage.

\subsection{Visualization} 

\subsubsection{Sentiment Analysis } 

\begin{figure}[!ht]
\centering
\includegraphics[width=3.8in]{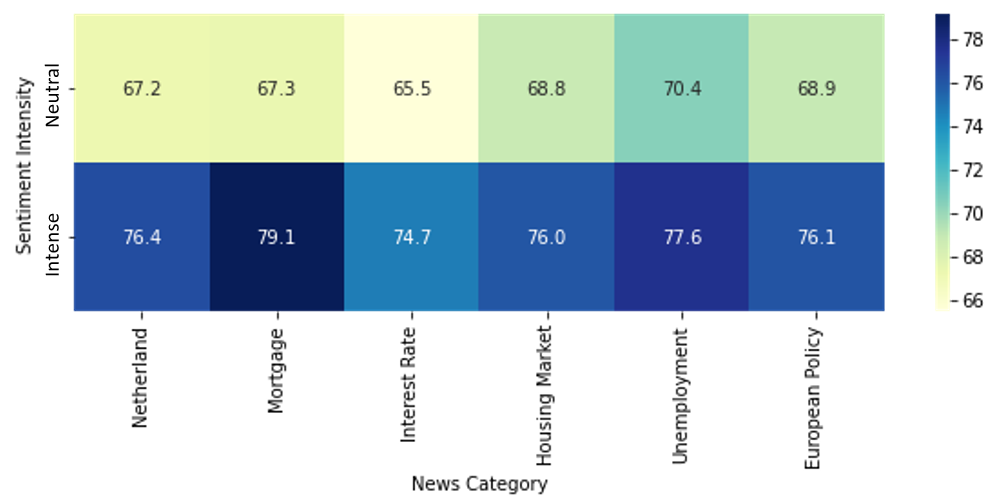}
\vspace{-0.3cm}
\caption{Heat map depicting the impact of sentiment intensity on accurately predicting mortgage risk based on text categories. 
}
\label{sentiment}
\end{figure}

Figure~\ref{sentiment} illustrates a heat map that represents the percentage of true predicted mortgage risk based on text categories and intensity of sentiment. The results suggest that the model's predictive precision experiences a notable increase as the sentiment score becomes more intense, surpassing an absolute threshold of 0.5. This implies that the influence of very positive or very negative news on decision-making regarding mortgage payments can result in a more accurate prediction of mortgage risk through the proposed model. This could be attributed to the fact that extremely positive or extremely negative news often elicits a more intense emotional response in individuals compared to neutral news.

Furthermore, our results imply that mortgage decisions are not purely rational; emotions may play a significant role. In other words, sentiment biases, such as optimism bias or pessimism bias, can also affect the' decision-making process of lenders \cite{koesoemasari2022investment}. If lenders are influenced by an optimism bias, they may be more inclined to approve loans even in situations where there may be underlying risks. In contrast, a pessimism bias can lead lenders to adopt a more conservative approach, resulting in potential missed lending opportunities.

\subsubsection{Model interpretation}

\begin{figure}[htbp!]
\centering
{\includegraphics[width=3.5in]{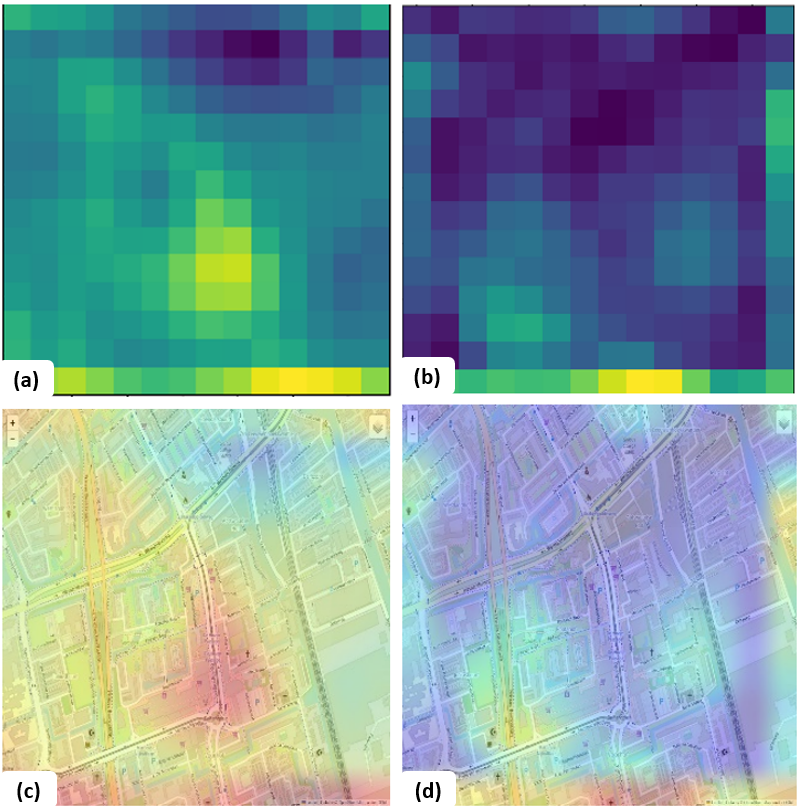}
\caption{Activation patterns in GradCAM: VGG vs. Multimodal approach for predicting high-risk mortgages. Panel (a) presents a heatmap generated by the VGG model emphasizing distinct color variations primarily driven by visual features.
Panel (b) presents a heatmap produced by the multimodal approach, showcasing less pronounced color variations due to the incorporation of additional information beyond images. Panel (c) presents  GradCAM visualization using the VGG model, highlighting significant activations in regions near main streets, indicative of low-risk mortgages. Panel (d) presents  GradCAM representation from the Multimodal approach, which is influenced by diverse data sources, including news and sentiment, demonstrating attenuated activation near main streets.}
\label{gradcam}}
\end{figure}

We extracted information from GradCAM using the VGG model (Figure ~\ref{gradcam}), which only used the image channel as input, and the resulting heatmap showed more distinct color variations compared to the multimodal approach. In GradCAM image data extracted from the multimodal model that features additional information beyond images, such as news and sentiment score (Figure ~\ref{gradcam} - panel d), there may still be activation areas in the same visual features as observed in the VGG model (panel c); however, the intensity of activation in those areas is much lower. 

The multimodal approach considers additional information beyond the image, such as the news and sentiment score. It seems that this broader input data modality introduces more complexity and potentially reduces the impact of the image's visual features on the final prediction. As a result, the color variations in the GradCAM heatmap generated by the multimodal approach are less differentiated compared to the VGG model. We notice that the proximity of areas to primary streets in the GradCAM visuals implies a substantial influence on the prediction mortgages with risks. Properties located near main streets are well known to offer several advantages, such as easy access to amenities, transportation infrastructure, and commercial centers.   The activation of areas close to main streets by the models indicates that they recognize visual features as indicators of properties in sought-after locations. Consequently, such properties are more likely to be associated with low-risk mortgages. In addition, main streets are typically found in well-established and stable neighborhoods to have better-maintained infrastructure and a sense of community. The main streets often serve as commercial hubs, attracting businesses and contributing to economic growth. Areas with strong local economies tend to have more stable housing markets and lower mortgage risk. 
Although we relied on qualitative Grad-CAM visualizations rather than quantitative interpretability metrics,   consistent alignment between Grad-CAM heatmaps and domain-relevant spatial features provides informal evidence that the model focuses on meaningful structures.

\section{Discussion}
\label{sec:discussion}
\subsection{Interpretation and Practical Implications}

Our empirical results demonstrate that unstructured information, including financial news media, sentiment cues, and LiDAR-derived spatial maps, can meaningfully approximate the predictive performance of traditional mortgage-risk models that rely on restricted proprietary, loan-level financial data. Our FusionCapsNet  with the capsule-inspired fusion mechanism  consistently outperformed common multimodal strategies such as Addition, Concatenation, and Cross-attention. Although absolute improvements in AUC and pAUC are moderate, they occur in a highly imbalanced setting and within the low–false-positive operating region most relevant to underwriting and portfolio-risk monitoring, and they are statistically significant, as our tests show. Even small gains in discrimination at low FPR can translate into fewer misclassified high-risk loans in practice, especially for institutions that must identify emerging distress early while minimizing unnecessary adverse decisions.

Unimodal analyzes clarify that each modality contributes a distinct and complementary perspective. Textual embeddings derived from BERT models capture macroeconomic narratives, regulatory developments, and sector-specific financial news that correlate with borrower stress or housing-market conditions. Our results indicate that LiDAR-based imagery provides fine-grained spatial signals—such as building density, infrastructure proximity, and neighbourhood form—that go far beyond what a postal code can express. The sentiment scores in the dataset provided an emotional-tone indicator that is not always evident in semantic content alone; the stronger predictive accuracy observed at extreme sentiment levels suggests that optimism or pessimism in news cycles meaningfully shapes lender and borrower behaviour(Figure~\ref{sentiment}).

The advantage of FusionCapsNet comes from its ability to preserve the structure of each modality through capsule projections, use agreement routing within each modality to filter inconsistent or noisy signals, and apply an adaptive gating mechanism to weight modalities based on their relevance to each prediction. This design is especially important in real-world situations where unstructured data quality varies widely—for example, when some regions have rich news coverage but a sparse LiDAR context and when sentiment becomes unusually polarized during macroeconomic volatility.

The contribution of sentiment scores as a feature in the multimodal setting aligns with the economic interpretation of mortgage risk. Negative changes in financial news tend to signal tightening economic conditions, such as increasing uncertainty in housing markets, interest-rate hikes, and increased unemployment concerns. These signals correlate with increased vulnerability of the borrower and greater likelihood of repayment challenges, thus raising the risk of default. In contrast, positive sentiment surrounding economic growth or policy support reduces risk perceptions and is reflected in lower predicted default probabilities. The improvement in performance of the model when sentiment features are included (Table~\ref{unimodal} and Table~\ref{performance}) indicates that such changes in the tone of the news offer early-warning signals that enhance the ability to detect high-risk situations emerging.

Beyond methodological contributions, these findings have important real-world implications,  FusionCapsNet can act as a cost-free early-warning system for credit unions, small lenders, housing agencies, and regulators who do not have access to proprietary scorecards. Unstructured data are globally available, inexpensive to collect, and continuously updated. Therefore, FusionCapsNet could support real-time monitoring of regional mortgage risk, highlighting areas where lenders may need to adjust underwriting standards, implement additional due diligence, or anticipate higher provisioning. In the case of regulators, multimodal scores can serve as a “shadow risk indicator” that complements internal models and indicates market-wide vulnerabilities that may not yet appear in supervisory datasets. For institutions with established underwriting frameworks, these unstructured data signals can serve as an auxiliary layer in model-risk management, supporting validation, trend detection, and comparison against internal risk assessments. The accompanying interpretability mechanisms—Grad-CAM for image-based reasoning and gating weights for modality contribution—further enhance the transparency of this approach, making it compatible with regulatory expectations regarding explainability.

\subsection{Limitations and Future Directions}

Despite its promising results, FusionCapsNet has limitations that point toward several avenues for future research. First, the model provides point estimates of default risk and does not quantify uncertainty. In real-world lending environments, uncertainty estimation is essential for understanding the degree of confidence in each prediction and preventing over-reliance on single outputs. Techniques such as Bayesian neural networks \cite{bhattacharya2019bayesian} or ensemble-based uncertainty quantification \cite{dastile2020statistical} could be integrated to provide calibrated risk intervals.

Second, although multiple visual backbones were evaluated under controlled experimental conditions, the present study does not benchmark the proposed fusion mechanism against more recent transformer-based visual architectures such as Vision Transformers (ViT) \cite{dosovitskiy2020image} or ConvNeXt\cite{liu2022convnet}. While VGG19 demonstrated the most stable performance for the current LiDAR-derived dataset, future research should systematically assess whether agreement-based capsule fusion yields complementary gains when paired with modern high-capacity encoders. Given that FusionCapsNet operates on modality-specific embeddings and is architecturally backbone-agnostic, integrating state-of-the-art visual encoders represents a natural and scalable extension.

Third, the sentiment channel, although informative, can capture short-lived reactions to news cycles that do not necessarily translate into long-term credit outcomes. Sudden bursts of pessimistic or optimistic reporting may influence predictions disproportionately without explicit temporal modeling. Future work could adopt a temporal  multimodal approach, where news flow, sentiment trajectories, and spatial signals are modeled longitudinally. Transformer-based sequence architectures or temporal graph models could capture lagged effects and structural breaks more effectively than the cross-sectional design used here.

Fourth, the empirical study focuses on Dutch mortgage portfolios due to the availability of high-quality LiDAR data and consistent loan records. Although this provides a strong starting point, the generalizability of the model across countries and housing markets remains untested. Extending the framework to other jurisdictions—especially those with different housing-market dynamics or less complete spatial data, would help assess its robustness and practical utility. This is especially relevant for developing countries, where spatial imagery and news sources may be available but high-resolution LiDAR may not.

Fifth, although FusionCapsNet can potentially be applied as a general multimodal fusion framework, it was primarily developed and evaluated for integrating textual, spatial, and sentiment-based information in the context of mortgage and credit-risk assessment.
The architecture was evaluated on a multimodal financial dataset that combines textual, spatial, and sentiment-based information relevant to credit-risk assessment. A limitation of the current work is the absence of experiments on standard public multimodal benchmarks.
However, commonly used financial datasets such as LendingClub are primarily structured and do not contain heterogeneous multimodal inputs, while widely used multimodal benchmarks from domains such as affective computing or vision-language learning differ substantially in both task objectives and data characteristics from mortgage-risk prediction. Consequently, direct comparison on these datasets may provide limited insight into performance within financial risk contexts. Future research will investigate the generalizability of FusionCapsNet across additional multimodal financial datasets as they become available and explore adaptations of the framework to other domains involving heterogeneous structured and unstructured data sources.

The framework could also benefit from richer or more diverse data sources. Integrating social-media signals, web-search trends, and additional macro-financial data could enhance predictive granularity, particularly for early detection of systemic stress. Techniques such as transfer learning and self-supervised learning\cite{weiss2016survey} would allow the model to leverage large corpora of unlabeled text or imagery, thereby strengthening representations before supervised training. For instance, pretraining the text encoder on broader macro-financial discourse or adapting the image encoder\cite{devlin2018bert} using urban-structure datasets may yield more robust embeddings for mortgage-risk prediction.

Also, improvements in explainability for the framework present a promising direction for future work. Although Grad-CAM offers insights into spatial drivers, emerging multimodal XAI approaches\cite{arrieta2020explainable} could provide unified explanations that connect textual cues, sentiment changes, and spatial features. Such methods would further align the framework with industry and regulatory expectations for transparent and auditable risk models. 

Finally, while Capsule Networks were originally developed for geometric part--whole modeling in vision, our use of capsules is confined to latent embedding space as an agreement-based aggregation mechanism. The model should therefore be interpreted as structured latent evidence alignment rather than geometric hierarchy modeling.

In summary, the study demonstrates that publicly accessible unstructured data can serve as a meaningful proxy for proprietary mortgage-risk indicators. FusionCapsNet’s capsule-inspired design effectively integrates heterogeneous information while preserving interpretability. Although more research is required to incorporate temporal dynamics, quantify uncertainty, expand geographic scope and enhance multimodal explainability, the findings support a broader shift towards alternative-data–driven risk assessment tools that are cost-effective, transparent, and suitable for a wide range of institutional contexts.

\section{Conclusion}
\label{sec:conclusion}

This study demonstrated that multimodal deep learning can enhance mortgage risk prediction by effectively utilizing unstructured information sources, such as financial news and geospatial imagery. The FusionCapsNet framework addressed the well-known limitations of early fusion strategies by preserving the modality-specific structure during integration, leading to more informative modality-level representations and adaptive integration of modality-level evidence summarized by normalized confidence scores through the gating mechanism. In our experiments, the fusion model consistently outperformed strong baselines, indicating that sentiment, textual narratives, and spatial neighborhood characteristics provide complementary risk signals that are not captured by structured loan data alone.

Our results show that beyond predictive gains, FusionCapsNet revealed practically relevant insights. We found that economic news sentiment amplified risk detection during unfavorable market conditions, while spatial features derived from LiDAR maps highlighted vulnerability associated with building density and exposure to environmental factors. The results suggest that unstructured data can serve as a timely early-warning layer for lenders and regulators, particularly when traditional borrower-level data is incomplete, proprietary, or stale. Grad-CAM interpretability further supports the operational value of the approach, as highlighted spatial patterns align with housing-market characteristics linked to mortgage distress. In this way, the model provides not only improved classification performance, but also transparency that can inform supervisory decision-making and risk-based reviews.

This work contributes an adaptable and scalable foundation for modern mortgage analytics by demonstrating that a fusion framework built around unstructured modalities can achieve performance comparable to commercial scorecards. The results motivate a deeper exploration of multimodal fusion in financial risk management, especially under shifting market conditions where traditional indicators may lag. Future research expanding temporal modeling, geographic scope, and quantitative interpretability analysis can advance this line of inquiry toward operational deployment.

\section{*Data and Code}
\label{sec:code}

The source code and data of the framework can be found in the following Github repository:
\\
\url{https://github.com/Banking-Analytics-Lab/UnstructuredMortgagePrediction}

 \section*{Acknowledgment}
The first and last authors acknowledge the support of the NSERC Discovery Grant program [RGPIN-2020-07114]. This research was undertaken in part thanks to funding from the Canada Research Chairs program [CRC-2024-00192]. This work was enabled in part by the support provided by Compute Ontario (\url{computeontario.ca}), Calcul Québec (\url{calculquebec.ca}), and the Digital Research Alliance of Canada (\url{alliancecan.ca}).


\bibliographystyle{elsarticle-num}
\bibliography{ref}

@article{bhattacharya2019bayesian,
  title={A Bayesian approach to modeling mortgage default and prepayment},
  author={Bhattacharya, Arnab and Wilson, Simon P and Soyer, Refik},
  journal={European Journal of Operational Research},
  volume={274},
  number={3},
  pages={1112--1124},
  year={2019},
  publisher={Elsevier}
}

@article{stevenson2021value,
  title={The value of text for small business default prediction: A deep learning approach},
  author={Stevenson, Matthew and Mues, Christophe and Bravo, Cristi{\'a}n},
  journal={European Journal of Operational Research},
  volume={295},
  number={2},
  pages={758--771},
  year={2021},
  publisher={Elsevier}
}

@article{dastile2020statistical,
  title={Statistical and machine learning models in credit scoring: A systematic literature survey},
  author={Dastile, Xolani and Celik, Turgay and Potsane, Moshe},
  journal={Applied Soft Computing},
  volume={91},
  pages={106263},
  year={2020},
  publisher={Elsevier}
}

@article{zhang2020combining,
  title={Combining structured and unstructured data for predictive models: a deep learning approach},
  author={Zhang, Dongdong and Yin, Changchang and Zeng, Jucheng and Yuan, Xiaohui and Zhang, Ping},
  journal={BMC medical informatics and decision making},
  volume={20},
  number={1},
  pages={1--11},
  year={2020},
  publisher={BioMed Central}
}

@article{feuerriegel2019news,
  title={News-based forecasts of macroeconomic indicators: A semantic path model for interpretable predictions},
  author={Feuerriegel, Stefan and Gordon, Julius},
  journal={European Journal of Operational Research},
  volume={272},
  number={1},
  pages={162--175},
  year={2019},
  publisher={Elsevier}
}

@article{devlin2018bert,
  title={Bert: Pre-training of deep bidirectional transformers for language understanding},
  author={Devlin, Jacob and Chang, Ming-Wei and Lee, Kenton and Toutanova, Kristina},
  journal={arXiv preprint arXiv:1810.04805},
  year={2018}
}

@article{ozturkkal2024explaining,
  title={Explaining mortgage defaults using SHAP and LASSO},
  author={Ozturkkal, Belma and Wahlstr{\o}m, Ranik Raaen},
  journal={Computational Economics},
  pages={1--35},
  year={2024},
  publisher={Springer}
}

@article{zheng2019multimodal,
  title={Multimodal deep network embedding with integrated structure and attribute information},
  author={Zheng, Conghui and Pan, Li and Wu, Peng},
  journal={IEEE transactions on neural networks and learning systems},
  volume={31},
  number={5},
  pages={1437--1449},
  year={2019},
  publisher={IEEE}
}

@article{cowden2019default,
  title={Default prediction of commercial real estate properties using machine learning techniques},
  author={Cowden, Chad and Fabozzi, Frank J and Nazemi, Abdolreza},
  journal={The Journal of Portfolio Management},
  volume={45},
  number={7},
  pages={55--67},
  year={2019},
  publisher={Institutional Investor Journals Umbrella}
}

@article{siering2023peer,
  title={Peer-to-peer (P2P) lending risk management: Assessing credit risk on social lending platforms using textual factors},
  author={Siering, Michael},
  journal={ACM Transactions on Management Information Systems},
  volume={14},
  number={3},
  pages={1--19},
  year={2023},
  publisher={ACM New York, NY}
}

@article{saavedra2024probability,
  title={Probability of default for lifetime credit loss for IFRS 9 using machine learning competing risks survival analysis models},
  author={Saavedra, Cayan Atreio Portela B{\'a}rcena and Fachini-Gomes, Juliana Betini and de Castro Gomes, Eduardo Monteiro and Kimura, Herbert},
  journal={Expert Systems with Applications},
  volume={249},
  pages={123607},
  year={2024},
  publisher={Elsevier}
}

@article{xia2020predicting,
  title={Predicting loan default in peer-to-peer lending using narrative data},
  author={Xia, Yufei and He, Lingyun and Li, Yinguo and Liu, Nana and Ding, Yanlin},
  journal={Journal of Forecasting},
  volume={39},
  number={2},
  pages={260--280},
  year={2020},
  publisher={Wiley Online Library}
}

@article{ge2017predicting,
  title={Predicting and deterring default with social media information in peer-to-peer lending},
  author={Ge, Ruyi and Feng, Juan and Gu, Bin and Zhang, Pengzhu},
  journal={Journal of Management Information Systems},
  volume={34},
  number={2},
  pages={401--424},
  year={2017},
  publisher={Taylor \& Francis}
}

@article{stevenson2022deep,
  title={Deep residential representations: Using unsupervised learning to unlock elevation data for geo-demographic prediction},
  author={Stevenson, Matthew and Mues, Christophe and Bravo, Cristi{\'a}n},
  journal={ISPRS Journal of Photogrammetry and Remote Sensing},
  volume={187},
  pages={378--392},
  year={2022},
  publisher={Elsevier}
}

@inproceedings{block2017unsupervised,
  title={An unsupervised deep learning approach for satellite image analysis with applications in demographic analysis},
  author={Block, Jessica and Yazdani, Mehrdad and Nguyen, Mai and Crawl, Daniel and Jankowska, Marta and Graham, John and DeFanti, Tom and Altintas, Ilkay},
  booktitle={2017 IEEE 13th International Conference on e-Science (e-Science)},
  pages={9--18},
  year={2017},
  organization={IEEE}
}

@article{jean2016combining,
  title={Combining satellite imagery and machine learning to predict poverty},
  author={Jean, Neal and Burke, Marshall and Xie, Michael and Davis, W Matthew and Lobell, David B and Ermon, Stefano},
  journal={Science},
  volume={353},
  number={6301},
  pages={790--794},
  year={2016},
  publisher={American Association for the Advancement of Science}
}

@article{law2019take,
  title={Take a look around: using street view and satellite images to estimate house prices},
  author={Law, Stephen and Paige, Brooks and Russell, Chris},
  journal={ACM Transactions on Intelligent Systems and Technology (TIST)},
  volume={10},
  number={5},
  pages={1--19},
  year={2019},
  publisher={ACM New York, NY, USA}
}

@article{zou2021detecting,
  title={Detecting individual abandoned houses from google street view: A hierarchical deep learning approach},
  author={Zou, Shengyuan and Wang, Le},
  journal={ISPRS Journal of Photogrammetry and Remote Sensing},
  volume={175},
  pages={298--310},
  year={2021},
  publisher={Elsevier}
}

@article{suel2021multimodal,
  title={Multimodal deep learning from satellite and street-level imagery for measuring income, overcrowding, and environmental deprivation in urban areas},
  author={Suel, Esra and Bhatt, Samir and Brauer, Michael and Flaxman, Seth and Ezzati, Majid},
  journal={Remote Sensing of Environment},
  volume={257},
  pages={112339},
  year={2021},
  publisher={Elsevier}
}

@article{pan2020land,
  title={Land-cover classification of multispectral LiDAR data using CNN with optimized hyper-parameters},
  author={Pan, Suoyan and Guan, Haiyan and Chen, Yating and Yu, Yongtao and Gon{\c{c}}alves, Wesley Nunes and Junior, Jos{\'e} Marcato and Li, Jonathan},
  journal={ISPRS Journal of Photogrammetry and Remote Sensing},
  volume={166},
  pages={241--254},
  year={2020},
  publisher={Elsevier}
}

@article{hamraz2019deep,
  title={Deep learning for conifer/deciduous classification of airborne LiDAR 3D point clouds representing individual trees},
  author={Hamraz, Hamid and Jacobs, Nathan B and Contreras, Marco A and Clark, Chase H},
  journal={ISPRS Journal of Photogrammetry and Remote Sensing},
  volume={158},
  pages={219--230},
  year={2019},
  publisher={Elsevier}
}

@article{zhou2020lidar,
  title={LiDAR-guided dense matching for detecting changes and updating of buildings in Airborne LiDAR data},
  author={Zhou, K and Lindenbergh, R and Gorte, Ben and Zlatanova, S},
  journal={ISPRS Journal of Photogrammetry and Remote Sensing},
  volume={162},
  pages={200--213},
  year={2020},
  publisher={Elsevier}
}

@article{lu2011volumetric,
  title={A volumetric approach to population estimation using LiDAR remote sensing},
  author={Lu, Zhenyu and Im, Jungho and Quackenbush, Lindi},
  journal={Photogrammetric Engineering \& Remote Sensing},
  volume={77},
  number={11},
  pages={1145--1156},
  year={2011},
  publisher={American Society for Photogrammetry and Remote Sensing}
}

@article{lu2013remote,
  title={Remote sensing-based house value estimation using an optimized regional regression model},
  author={Lu, Zhenyu and Im, Jungho and Quackenbush, Lindi J and Yoo, Sanglim},
  journal={Photogrammetric Engineering \& Remote Sensing},
  volume={79},
  number={9},
  pages={809--820},
  year={2013},
  publisher={American Society for Photogrammetry and Remote Sensing}
}

@article{grove2014ecology,
  title={An ecology of prestige in New York City: Examining the relationships among population density, socio-economic status, group identity, and residential canopy cover},
  author={Grove, J Morgan and Locke, Dexter H and O’Neil-Dunne, Jarlath PM},
  journal={Environmental management},
  volume={54},
  pages={402--419},
  year={2014},
  publisher={Springer}
}

@article{shanahan2014socio,
  title={Socio-economic inequalities in access to nature on public and private lands: A case study from Brisbane, Australia},
  author={Shanahan, Danielle F and Lin, BB and Gaston, KJ and Bush, R and Fuller, RA},
  journal={Landscape and urban planning},
  volume={130},
  pages={14--23},
  year={2014},
  publisher={Elsevier}
}

@article{warth2020prediction,
  title={Prediction of socio-economic indicators for urban planning using VHR satellite imagery and spatial analysis},
  author={Warth, Gebhard and Braun, Andreas and Assmann, Oliver and Fleckenstein, Kevin and Hochschild, Volker},
  journal={Remote Sensing},
  volume={12},
  number={11},
  pages={1730},
  year={2020},
  publisher={Mdpi}
}

@article{li2017web,
  title={Web media and stock markets: A survey and future directions from a big data perspective},
  author={Li, Qing and Chen, Yan and Wang, Jun and Chen, Yuanzhu and Chen, Hsinchun},
  journal={IEEE Transactions on Knowledge and Data Engineering},
  volume={30},
  number={2},
  pages={381--399},
  year={2017},
  publisher={IEEE}
}

@article{farimani2022investigating,
  title={Investigating the informativeness of technical indicators and news sentiment in financial market price prediction},
  author={Farimani, Saeede Anbaee and Jahan, Majid Vafaei and Fard, Amin Milani and Tabbakh, Seyed Reza Kamel},
  journal={Knowledge-Based Systems},
  volume={247},
  pages={108742},
  year={2022},
  publisher={Elsevier}
}

@article{chen2022comparative,
  title={A comparative study of automated legal text classification using random forests and deep learning},
  author={Chen, Haihua and Wu, Lei and Chen, Jiangping and Lu, Wei and Ding, Junhua},
  journal={Information Processing \& Management},
  volume={59},
  number={2},
  pages={102798},
  year={2022},
  publisher={Elsevier}
}

@article{minaee2021deep,
  title={Deep learning--based text classification: a comprehensive review},
  author={Minaee, Shervin and Kalchbrenner, Nal and Cambria, Erik and Nikzad, Narjes and Chenaghlu, Meysam and Gao, Jianfeng},
  journal={ACM computing surveys (CSUR)},
  volume={54},
  number={3},
  pages={1--40},
  year={2021},
  publisher={ACM New York, NY, USA}
}

@article{nasir2021fake,
  title={Fake news detection: A hybrid CNN-RNN based deep learning approach},
  author={Nasir, Jamal Abdul and Khan, Osama Subhani and Varlamis, Iraklis},
  journal={International Journal of Information Management Data Insights},
  volume={1},
  number={1},
  pages={100007},
  year={2021},
  publisher={Elsevier}
}

@article{liao2021integrated,
  title={An integrated multi-task model for fake news detection},
  author={Liao, Qing and Chai, Heyan and Han, Hao and Zhang, Xiang and Wang, Xuan and Xia, Wen and Ding, Ye},
  journal={IEEE Transactions on Knowledge and Data Engineering},
  volume={34},
  number={11},
  pages={5154--5165},
  year={2021},
  publisher={IEEE}
}

@article{hu2022causal,
  title={Causal Inference for Leveraging Image-text Matching Bias in Multi-modal Fake News Detection},
  author={Hu, Linmei and Chen, Ziwei and Yin, Ziwang Zhao Jianhua and Nie, Liqiang},
  journal={IEEE Transactions on Knowledge and Data Engineering},
  year={2022},
  publisher={IEEE}
}

@article{li2020multimodal,
  title={A multimodal event-driven lstm model for stock prediction using online news},
  author={Li, Qing and Tan, Jinghua and Wang, Jun and Chen, Hsinchun},
  journal={IEEE Transactions on Knowledge and Data Engineering},
  volume={33},
  number={10},
  pages={3323--3337},
  year={2020},
  publisher={IEEE}
}

@article{ramisa2017breakingnews,
  title={Breakingnews: Article annotation by image and text processing},
  author={Ramisa, Arnau and Yan, Fei and Moreno-Noguer, Francesc and Mikolajczyk, Krystian},
  journal={IEEE transactions on pattern analysis and machine intelligence},
  volume={40},
  number={5},
  pages={1072--1085},
  year={2017},
  publisher={IEEE}
}

@article{yang2019shared,
  title={Shared multi-view data representation for multi-domain event detection},
  author={Yang, Zhenguo and Li, Qing and Liu, Wenyin and Lv, Jianming},
  journal={IEEE transactions on pattern analysis and machine intelligence},
  volume={42},
  number={5},
  pages={1243--1256},
  year={2019},
  publisher={IEEE}
}

@article{isnan2023sentiment,
  title={Sentiment analysis for TikTok review using VADER sentiment and SVM model},
  author={Isnan, Mahmud and Elwirehardja, Gregorius Natanael and Pardamean, Bens},
  journal={Procedia Computer Science},
  volume={227},
  pages={168--175},
  year={2023},
  publisher={Elsevier}
}

@article{liu2019roberta,
  title={Roberta: A robustly optimized bert pretraining approach},
  author={Liu, Yinhan and Ott, Myle and Goyal, Naman and Du, Jingfei and Joshi, Mandar and Chen, Danqi and Levy, Omer and Lewis, Mike and Zettlemoyer, Luke and Stoyanov, Veselin},
  journal={arXiv preprint arXiv:1907.11692},
  year={2019}
}

@article{lan2019albert,
  title={Albert: A lite bert for self-supervised learning of language representations},
  author={Lan, Zhenzhong and Chen, Mingda and Goodman, Sebastian and Gimpel, Kevin and Sharma, Piyush and Soricut, Radu},
  journal={arXiv preprint arXiv:1909.11942},
  year={2019}
}

@inproceedings{tan2019efficientnet,
  title={Efficientnet: Rethinking model scaling for convolutional neural networks},
  author={Tan, Mingxing and Le, Quoc},
  booktitle={International conference on machine learning},
  pages={6105--6114},
  year={2019},
  organization={PMLR}
}

@article{chen2014big,
  title={Big data deep learning: challenges and perspectives},
  author={Chen, Xue-Wen and Lin, Xiaotong},
  journal={IEEE access},
  volume={2},
  pages={514--525},
  year={2014},
  publisher={Ieee}
}

@article{jiang2018loan,
  title={Loan default prediction by combining soft information extracted from descriptive text in online peer-to-peer lending},
  author={Jiang, Cuiqing and Wang, Zhao and Wang, Ruiya and Ding, Yong},
  journal={Annals of Operations Research},
  volume={266},
  number={1-2},
  pages={511--529},
  year={2018},
  publisher={Springer}
}

@article{adrian2021sentinel,
  title={Sentinel SAR-optical fusion for crop type mapping using deep learning and Google Earth Engine},
  author={Adrian, Jarrett and Sagan, Vasit and Maimaitijiang, Maitiniyazi},
  journal={ISPRS Journal of Photogrammetry and Remote Sensing},
  volume={175},
  pages={215--235},
  year={2021},
  publisher={Elsevier}
}

@article{zhu2017housing,
  title={Housing market stability, mortgage market structure, and monetary policy: Evidence from the euro area},
  author={Zhu, Bing and Betzinger, Michael and Sebastian, Steffen},
  journal={Journal of Housing Economics},
  volume={37},
  pages={1--21},
  year={2017},
  publisher={Elsevier}
}

@phdthesis{iselin2016news,
  title={News and the Economy--How to Measure Economic Trends by Using Media-Based Data},
  author={Iselin, David},
  year={2016},
  school={ETH Zurich}
}

@article{vaswani2017attention,
  title={Attention is all you need},
  author={Vaswani, Ashish and Shazeer, Noam and Parmar, Niki and Uszkoreit, Jakob and Jones, Llion and Gomez, Aidan N and Kaiser, {\L}ukasz and Polosukhin, Illia},
  journal={Advances in neural information processing systems},
  volume={30},
  year={2017}
}

@article{zhao2024deep,
  title={Deep multimodal data fusion},
  author={Zhao, Fei and Zhang, Chengcui and Geng, Baocheng},
  journal={ACM computing surveys},
  volume={56},
  number={9},
  pages={1--36},
  year={2024},
  publisher={ACM New York, NY}
}

@article{pawlowski2023effective,
  title={Effective techniques for multimodal data fusion: A comparative analysis},
  author={Paw{\l}owski, Maciej and Wr{\'o}blewska, Anna and Sysko-Roma{\'n}czuk, Sylwia},
  journal={Sensors},
  volume={23},
  number={5},
  pages={2381},
  year={2023},
  publisher={MDPI}
}

@article{kealhofer2003quantifying,
  author    = {Stephen Kealhofer},
  title     = {Quantifying Credit Risk I: Default Prediction},
  journal   = {Financial Analysts Journal},
  volume    = {59},
  number    = {1},
  year      = {2003},
  pages     = {30--44},
}

@article{chen2002approach,
  title={An approach to linking remotely sensed data and areal census data},
  author={Chen, K},
  journal={International Journal of Remote Sensing},
  volume={23},
  number={1},
  pages={37--48},
  year={2002},
  publisher={Taylor \& Francis}
}

@article{li2007measuring,
  title={Measuring the quality of life in city of Indianapolis by integration of remote sensing and census data},
  author={Li, G and Weng, Qiaho},
  journal={International Journal of Remote Sensing},
  volume={28},
  number={2},
  pages={249--267},
  year={2007},
  publisher={Taylor \& Francis}
}

@article{SHUKLA2023100025,
    title = {An evaluation of Google Translate for Sanskrit to English translation via sentiment and semantic analysis},
    journal = {Natural Language Processing Journal},
    volume = {4},
    pages = {100025},
    year = {2023},
    issn = {2949-7191},
  author={Shukla, Akshat and Bansal, Chaarvi and Badhe, Sushrut and Ranjan, Mukul and Chandra, Rohitash}
}

@article{li2024geotpe,
  title={GeoTPE: A neural network model for geographical topic phrases extraction from literature based on BERT enhanced with relative position embedding},
  author={Li, Weirong and Sun, Kai and Zhu, Yunqiang and Ding, Fangyu and Hu, Lei and Dai, Xiaoliang and Song, Jia and Yang, Jie and Qian, Lang and Wang, Shu},
  journal={Expert Systems with Applications},
  volume={235},
  pages={121077},
  year={2024},
  publisher={Elsevier}
}

@article{liu2023naming,
  title={Naming entity recognition of citrus pests and diseases based on the BERT-BiLSTM-CRF model},
  author={Liu, Yafei and Wei, Siqi and Huang, Haijun and Lai, Qin and Li, Mengshan and Guan, Lixin},
  journal={Expert Systems with Applications},
  volume={234},
  pages={121103},
  year={2023},
  publisher={Elsevier}
}

@article{briskilal2022ensemble,
  title={An ensemble model for classifying idioms and literal texts using BERT and RoBERTa},
  author={Briskilal, J and Subalalitha, CN},
  journal={Information Processing \& Management},
  volume={59},
  number={1},
  pages={102756},
  year={2022},
  publisher={Elsevier}
}

@article{wu2021research,
  title={Research on named entity recognition of electronic medical records based on roberta and radical-level feature},
  author={Wu, Yue and Huang, Jie and Xu, Caie and Zheng, Huilin and Zhang, Lei and Wan, Jian},
  journal={Wireless Communications and Mobile Computing},
  volume={2021},
  pages={1--10},
  year={2021},
  publisher={Hindawi Limited}
}

@article{bansal2023transfer,
  title={Transfer learning for image classification using VGG19: Caltech-101 image data set},
  author={Bansal, Monika and Kumar, Munish and Sachdeva, Monika and Mittal, Ajay},
  journal={Journal of ambient intelligence and humanized computing},
  volume={14},
  number={4},
  pages={3609--3620},
  year={2023},
  publisher={Springer}
}

@article{bansal2021transfer,
  title={Transfer learning for image classification using VGG19: Caltech-101 image data set},
  author={Bansal, Monika and Kumar, Munish and Sachdeva, Monika and Mittal, Ajay},
  journal={Journal of ambient intelligence and humanized computing},
  pages={1--12},
  year={2021},
  publisher={Springer}
}

@article{elsken2019neural,
  title={Neural architecture search: A survey},
  author={Elsken, Thomas and Metzen, Jan Hendrik and Hutter, Frank},
  journal={The Journal of Machine Learning Research},
  volume={20},
  number={1},
  pages={1997--2017},
  year={2019},
  publisher={JMLR. org}
}

@inproceedings{he2016deep,
  title={Deep residual learning for image recognition},
  author={He, Kaiming and Zhang, Xiangyu and Ren, Shaoqing and Sun, Jian},
  booktitle={Proceedings of the IEEE conference on computer vision and pattern recognition},
  pages={770--778},
  year={2016}
}

@article{zhao2019object,
  title={Object detection with deep learning: A review},
  author={Zhao, Zhong-Qiu and Zheng, Peng and Xu, Shou-tao and Wu, Xindong},
  journal={IEEE transactions on neural networks and learning systems},
  volume={30},
  number={11},
  pages={3212--3232},
  year={2019},
  publisher={IEEE}
}

@article{bajpai2025ri,
  title={RI-L1Approx: A novel Resnet-Inception-based Fast L1-approximation method for face recognition},
  author={Bajpai, Supriya and Mishra, Gargi and Jain, Rachna and Jain, Deepak Kumar and Saini, Dharmender and Hussain, Amir},
  journal={Neurocomputing},
  volume={613},
  pages={128708},
  year={2025},
  publisher={Elsevier}
}

@article{gill2023health,
  title={Health uninsurance premium and mortgage interest rates},
  author={Gill, Balbinder Singh},
  journal={International Review of Financial Analysis},
  volume={87},
  pages={102647},
  year={2023},
  publisher={Elsevier}
}

@article{magri2011rise,
  title={The rise of risk-based pricing of mortgage interest rates in Italy},
  author={Magri, Silvia and Pico, Raffaella},
  journal={Journal of Banking \& Finance},
  volume={35},
  number={5},
  pages={1277--1290},
  year={2011},
  publisher={Elsevier}
}

@article{hodge2017assessment,
  title={Assessment inequity in a declining housing market: The case of Detroit},
  author={Hodge, Timothy R and McMillen, Daniel P and Sands, Gary and Skidmore, Mark},
  journal={Real Estate Economics},
  volume={45},
  number={2},
  pages={237--258},
  year={2017},
  publisher={Wiley Online Library}
}

@article{gabriel1991credit,
  title={Credit rationing, race, and the mortgage market},
  author={Gabriel, Stuart A and Rosenthal, Stuart S},
  journal={Journal of Urban Economics},
  volume={29},
  number={3},
  pages={371--379},
  year={1991},
  publisher={Elsevier}
}

@article{gyourko2014reconciling,
  title={Reconciling theory and empirics on the role of unemployment in mortgage default},
  author={Gyourko, Joseph and Tracy, Joseph},
  journal={Journal of Urban Economics},
  volume={80},
  pages={87--96},
  year={2014},
  publisher={Elsevier}
}

@article{mak2015responsible,
  title={What is responsible lending? The EU consumer mortgage credit directive in the UK and the Netherlands},
  author={Mak, Vanessa},
  journal={Journal of Consumer Policy},
  volume={38},
  pages={411--430},
  year={2015},
  publisher={Springer}
}

@article{koesoemasari2022investment,
  title={Investment Strategy Based on Bias Behavior and Investor Sentiment in Emerging Markets},
  author={Koesoemasari, Dian Safitri Pantja and Haryono, Tulus and Trinugroho, Irwan and Setiawan, Doddy},
  journal={Etikonomi},
  volume={21},
  number={1},
  pages={1--10},
  year={2022}
}

@article{weiss2016survey,
  title={A survey of transfer learning},
  author={Weiss, Karl and Khoshgoftaar, Taghi M and Wang, DingDing},
  journal={Journal of Big data},
  volume={3},
  number={1},
  pages={1--40},
  year={2016},
  publisher={SpringerOpen}
}

@article{arrieta2020explainable,
  title={Explainable Artificial Intelligence (XAI): Concepts, taxonomies, opportunities and challenges toward responsible AI},
  author={Arrieta, Alejandro Barredo and D{\'\i}az-Rodr{\'\i}guez, Natalia and Del Ser, Javier and Bennetot, Adrien and Tabik, Siham and Barbado, Alberto and Garc{\'\i}a, Salvador and Gil-L{\'o}pez, Sergio and Molina, Daniel and Benjamins, Richard and others},
  journal={Information fusion},
  volume={58},
  pages={82--115},
  year={2020},
  publisher={Elsevier}
}

@inproceedings{shayaa2017social,
  title={Social media sentiment analysis on employment in Malaysia},
  author={Shayaa, Shahid and Wai, Phoong Seuk and Chung, Yeong Wai and Sulaiman, Ainin and Jaafar, Noor Ismawati and Zakaria, Shamshul Bahri},
  booktitle={the Proceedings of 8th Global Business and Finance Research Conference, Taipei, Taiwan},
  year={2017}
}

@article{gao2023severe,
  title={Severe weather and peer-to-peer farmers’ loan default predictions: Evidence from machine learning analysis},
  author={Gao, Wei and Ju, Ming and Yang, Tongyang},
  journal={Finance Research Letters},
  volume={58},
  pages={104287},
  year={2023},
  publisher={Elsevier}
}

@article{zheng2023community,
  title={Community resilience and house prices: A machine learning approach},
  author={Zheng, Yi},
  journal={Finance Research Letters},
  volume={58},
  pages={104400},
  year={2023},
  publisher={Elsevier}
}

@article{nwafor2023determinants,
  title={Determinants of non-performing loans: An explainable ensemble and deep neural network approach},
  author={Nwafor, Chioma Ngozi and Nwafor, Obumneme Zimuzor},
  journal={Finance Research Letters},
  volume={56},
  pages={104084},
  year={2023},
  publisher={Elsevier}
}

@article{gupta2018digital,
  title={Digital elevation model},
  author={Gupta, Ravi P and Gupta, Ravi P},
  journal={Remote Sensing Geology},
  pages={101--106},
  year={2018},
  publisher={Springer}
}

@article{li2022brief,
  title={A brief overview of universal sentence representation methods: A linguistic view},
  author={Li, Ruiqi and Zhao, Xiang and Moens, Marie-Francine},
  journal={ACM Computing Surveys (CSUR)},
  volume={55},
  number={3},
  pages={1--42},
  year={2022},
  publisher={ACM New York, NY}
}

@inproceedings{gulshan2019effnet,
  title={Performance of a deep-learning algorithm vs manual grading for detecting diabetic retinopathy in India},
  author={Gulshan, Varun and Rajan, Ramasamy P and Widner, Kim and Wu, David and Wubbels, Patrick and Rhodes, Taylor and Whitehouse, Claire R and Coram, Marc and Corrado, Greg S and Ramasamy, Krishnakumar and Raman, Rajiv and others},
  booktitle={JAMA ophthalmology},
  volume={137},
  number={9},
  pages={987--993},
  year={2019},
  organization={American Medical Association}
}

@article{ferentinos2018deep,
  title={Deep learning models for plant disease detection and diagnosis},
  author={Ferentinos, Konstantinos P},
  journal={Computers and Electronics in Agriculture},
  volume={145},
  pages={311--318},
  year={2018},
  publisher={Elsevier}
}

@article{todd2025multimodal,
  title={A multimodal sentiment classifier for financial decision making},
  author={Todd, Andrew and Bowden, James and Cummins, Mark and Su, Yang},
  journal={International Review of Financial Analysis},
  pages={104322},
  year={2025},
  publisher={Elsevier}
}

@article{wang2023attentive,
  title={Attentive statement fraud detection: Distinguishing multimodal financial data with fine-grained attention},
  author={Wang, Gang and Ma, Jingling and Chen, Gang},
  journal={Decision Support Systems},
  volume={167},
  pages={113913},
  year={2023},
  publisher={Elsevier}
}

@article{peng2025multimodal,
  title={Multimodal fusion hybrid neural network approach for multi-class damage classification in high-speed rail track-bridge systems with multi-parameter},
  author={Peng, Kang and Zhou, Wangbao and Jiang, Lizhong and Xiong, Lijun and Yan, Wang-Ji},
  journal={Engineering Structures},
  volume={328},
  pages={119710},
  year={2025},
  publisher={Elsevier}
}

@article{jiao2024comprehensive,
  title={A Comprehensive Survey on Deep Learning Multi-Modal Fusion: Methods, Technologies and Applications.},
  author={Jiao, Tianzhe and Guo, Chaopeng and Feng, Xiaoyue and Chen, Yuming and Song, Jie},
  journal={Computers, Materials \& Continua},
  volume={80},
  number={1},
  year={2024}
}

@inproceedings{gaonkar2021comprehensive,
  title={A comprehensive survey on multimodal data representation and information fusion algorithms},
  author={Gaonkar, Apeksha and Chukkapalli, Yogya and Raman, P Jahnavi and Srikanth, Sahana and Gurugopinath, Sanjeev},
  booktitle={2021 international conference on intelligent technologies (CONIT)},
  pages={1--8},
  year={2021},
  organization={IEEE}
}

@article{gan2024multimodal,
  title={A multimodal fusion network with attention mechanisms for visual--textual sentiment analysis},
  author={Gan, Chenquan and Fu, Xiang and Feng, Qingdong and Zhu, Qingyi and Cao, Yang and Zhu, Ye},
  journal={Expert Systems with Applications},
  volume={242},
  pages={122731},
  year={2024},
  publisher={Elsevier}
}

@article{wang2025cross,
  title={Cross-attention interaction learning network for multi-model image fusion via transformer},
  author={Wang, Jing and Yu, Long and Tian, Shengwei},
  journal={Engineering Applications of Artificial Intelligence},
  volume={139},
  pages={109583},
  year={2025},
  publisher={Elsevier}
}

@inproceedings{an2024attention,
  title={Attention-based Multimodal Fusion with Adversarial Network for In-Hospital Mortality Prediction},
  author={An, Ying and Qiu, Ruping and Guo, Lin and Chen, Xianlai},
  booktitle={2024 IEEE International Conference on Bioinformatics and Biomedicine (BIBM)},
  pages={1789--1795},
  year={2024},
  organization={IEEE}
}

@article{li2024crossfuse,
  title={CrossFuse: A novel cross attention mechanism based infrared and visible image fusion approach},
  author={Li, Hui and Wu, Xiao-Jun},
  journal={Information Fusion},
  volume={103},
  pages={102147},
  year={2024},
  publisher={Elsevier}
}

@article{duan2024hybrid,
  title={A hybrid neural network model for sentiment analysis of financial texts using topic extraction, pre-trained model, and enhanced attention mechanism methods},
  author={Duan, Ganglong and Yan, Shunfei and Zhang, Meng},
  journal={IEEE Access},
  year={2024},
  publisher={IEEE}
}

@article{vaca2024interpretability,
  title={Interpretability of deep learning models in analysis of Spanish financial text},
  author={Vaca, C{\'e}sar and Astorgano, Manuel and Lopez-Rivero, Alfonso J and Tejerina, Fernando and Sahelices, Benjam{\'\i}n},
  journal={Neural Computing and Applications},
  volume={36},
  number={13},
  pages={7509--7527},
  year={2024},
  publisher={Springer}
}

@article{shishehgarkhaneh2024transformer,
  title={Transformer-based named entity recognition in construction supply chain risk management in Australia},
  author={Shishehgarkhaneh, Milad Baghalzadeh and Moehler, Robert C and Fang, Yihai and Hijazi, Amer A and Aboutorab, Hamed},
  journal={IEEE Access},
  volume={12},
  pages={41829--41851},
  year={2024},
  publisher={IEEE}
}

@article{thekkekara2024attention,
  title={An attention-based CNN-BiLSTM model for depression detection on social media text},
  author={Thekkekara, Joel Philip and Yongchareon, Sira and Liesaputra, Veronica},
  journal={Expert systems with applications},
  volume={249},
  pages={123834},
  year={2024},
  publisher={Elsevier}
}

@article{sabour2017dynamic,
  title={Dynamic routing between capsules},
  author={Sabour, Sara and Frosst, Nicholas and Hinton, Geoffrey E},
  journal={Advances in neural information processing systems},
  volume={30},
  year={2017}
}

@article{lalonde2021capsules,
  title={Capsules for biomedical image segmentation},
  author={LaLonde, Rodney and Xu, Ziyue and Irmakci, Ismail and Jain, Sanjay and Bagci, Ulas},
  journal={Medical image analysis},
  volume={68},
  pages={101889},
  year={2021},
  publisher={Elsevier}
}

@article{paoletti2018capsule,
  title={Capsule networks for hyperspectral image classification},
  author={Paoletti, Mercedes E and Haut, Juan Mario and Fernandez-Beltran, Ruben and Plaza, Javier and Plaza, Antonio and Li, Jun and Pla, Filiberto},
  journal={IEEE Transactions on Geoscience and Remote Sensing},
  volume={57},
  number={4},
  pages={2145--2160},
  year={2018},
  publisher={IEEE}
}

@article{kim2020text,
  title={Text classification using capsules},
  author={Kim, Jaeyoung and Jang, Sion and Park, Eunjeong and Choi, Sungchul},
  journal={Neurocomputing},
  volume={376},
  pages={214--221},
  year={2020},
  publisher={Elsevier}
}

@article{ntiamoah2014loan,
  title={Loan default rate and its impact on profitability in financial institutions},
  author={Ntiamoah, Evans Brako and Oteng, Emmanuel and Opoku, Beatrice and Siaw, Anthony},
  journal={Research journal of Finance and accounting},
  volume={5},
  number={14},
  pages={67--72},
  year={2014}
}

@article{tavakoli2025multi,
  title={Multi-modal deep learning for credit rating prediction using text and numerical data streams},
  author={Tavakoli, Mahsa and Chandra, Rohitash and Tian, Fengrui and Bravo, Cristi{\'a}n},
  journal={Applied Soft Computing},
  volume={171},
  pages={112771},
  year={2025},
  publisher={Elsevier}
}

@misc{EUDW2022,
  author       = {{European DataWarehouse}},
  title        = {Residential Mortgage Data for the Netherlands},
  year         = {2022},
  howpublished = {\url{https://editor.eurodw.eu/}},
  note         = {Accessed 2023-05-01}
}

@misc{AHN2_2022,
  author       = {{Actueel Hoogtebestand Nederland (AHN)}},
  title        = {AHN2 0.5m Resolution Digital Elevation Model (DEM)},
  year         = {2022},
  howpublished = {\url{https://developers.google.com/earth-engine/datasets/catalog/AHN_AHN2_05M_RUW}},
  note         = {Accessed 2023-05-01}
}

@misc{seekingalpha,
  title        = {Seeking Alpha Market News},
  howpublished = {\url{https://seekingalpha.com/market-news}},
  note         = {Accessed 2024}
}

@misc{earthenginecatalog,
  title        = {Google Earth Engine Data Catalog},
  howpublished = {\url{https://developers.google.com/earth-engine/datasets/catalog}},
  note         = {Accessed 2024}
}

@misc{google_earth_engine,
  title        = {{Google Earth Engine}},
  author       = {{Google LLC}},
  year         = {2024},
  howpublished = {\url{https://earthengine.google.com/}},
  note         = {Accessed: 2025-01-15}
}

@article{dosovitskiy2020image,
  title={An image is worth 16x16 words: Transformers for image recognition at scale},
  author={Dosovitskiy, Alexey and Beyer, Lucas and Kolesnikov, Alexander and Weissenborn, Dirk and Zhai, Xiaohua and Unterthiner, Thomas and Dehghani, Mostafa and Minderer, Matthias and Heigold, Georg and Gelly, Sylvain and others},
  journal={arXiv preprint arXiv:2010.11929},
  year={2020}
}

@inproceedings{liu2022convnet,
  title={A convnet for the 2020s},
  author={Liu, Zhuang and Mao, Hanzi and Wu, Chao-Yuan and Feichtenhofer, Christoph and Darrell, Trevor and Xie, Saining},
  booktitle={Proceedings of the IEEE/CVF conference on computer vision and pattern recognition},
  pages={11976--11986},
  year={2022}
}

\end{document}